\documentclass[12pt]{article}

\usepackage{cite}
\usepackage{subfigure}
\usepackage{multirow}
\usepackage{helvet}
\usepackage{amsmath}
\usepackage{amssymb}
\usepackage{setspace}
\usepackage{setspace}
\usepackage{graphicx}
\usepackage{empheq}
\usepackage{subfigure}
\usepackage{wrapfig}
\usepackage{float}
\usepackage{cleveref}
\usepackage{longtable}
\usepackage{color}

\def\be{\begin{equation}}
\def\ee{\end{equation}}

\begin{document}

\titlepage                                                    

 \vspace*{0.5cm}
\begin{center}                                                    
{\Large \bf Ad Lucem: QED Parton Distribution \\ \vspace{0.3cm} Functions in the MMHT Framework}\\

\vspace*{1cm}
                                                   
L. A. Harland--Lang$^{1}$, A. D. Martin$^{2}$, R. Nathvani$^{3}$, R. S. Thorne$^{3}$. \\                                                 
                                                   
\vspace*{0.5cm}
${}^1$Rudolf Peierls Centre, Beecroft Building, Parks Road, Oxford, OX1 3PU\\                                                   
${}^2$Institute for Particle Physics Phenomenology, University of Durham, Durham, DH1 3LE \\                                      
${}^3$Department of Physics and Astronomy, University College London, WC1E 6BT, UK 
                                                    
\vspace*{1cm}                         

\begin{abstract}

\noindent  We present the MMHT2015qed PDF set,  resulting from the inclusion of QED corrections to the existing set of MMHT Parton Distribution Functions (PDFs), and which contain the photon PDF of the proton. Adopting an input distribution from the LUXqed formulation, we discuss our methods of including QED effects for the full, coupled DGLAP evolution of all partons with QED at $\mathcal{O}(\alpha)$, $\mathcal{O}(\alpha\alpha_{S})$, $\mathcal{O}(\alpha^2)$. While we find consistency for the photon PDF of the proton with other recent sets, building on this we also present a set of QED corrected neutron PDFs and provide the photon PDF separated into its elastic and inelastic contributions. The effect of QED corrections on the other partons and the fit quality is investigated, and the sources of uncertainty for the photon are outlined. Finally we explore the phenomenological implications of this set, giving the partonic luminosities for both the elastic and inelastic contributions to the photon and the effect of our photon PDF on fits to high mass Drell-Yan production, including the photon--initiated channel.

\end{abstract}
                                   
\end{center}

\tableofcontents

\section{Introduction}\label{sec_intro}

The precision physics program at the Large Hadron Collider (LHC) aims to observe processes at an unprecedented level of accuracy and experimental sensitivity. 
As part of these efforts, the analyses conducted by the LHC experimental collaborations are increasingly undertaken with theoretical cross section predictions at next--to--next--to--leading order (NNLO) in QCD, which includes $\mathcal{O}(\alpha_S^2)$ corrections. 
At this level of precision, it is expected that electroweak (EW) corrections, including those with photon--initiated (PI) processes, will begin to have observable effects as $\alpha_{QED} \sim \alpha_{S}^2$ at the typical scales being probed at the LHC. These should therefore be incorporated in theoretical predictions. In particular, electroweak corrected partonic cross sections should be calculated with corresponding Parton Distribution Functions (PDFs)
produced at NLO and NNLO in QCD and the appropriate order in QED.
This is achieved primarily by modifying the DGLAP~\cite{DGLAP1,DGLAP2,DGLAP3} factorisation scale evolution of the PDFs to include QED parton splittings.
The most significant effect of this change is the necessary inclusion of the photon as a constituent parton of the proton. Subsequently one can also begin to calculate the effect of PI sub-processes as corrections to the leading QCD cross section for processes such as Drell--Yan \cite{DY-ATLAS}, EW boson--boson scattering \cite{W_W} and Higgs production with an associated EW boson \cite{Higgs+W}, which are expected to be sensitive to these effects. 
In a different context, semi-exclusive \cite{Harland-Lang:2016apc} and exclusive production of states with EW couplings are also related to the photon content of the proton, albeit not directly to the inclusive photon PDF.  Here, PI processes play an important role, see e.g.~\cite{Harland-Lang:2018hmi,Beresford:2018pbt} for recent studies in the context of compressed SUSY scenarios.

MRST provided the first such publicly available QED set\cite{mrstqed}, modelling the photon at the input scale as arising radiatively from the quarks (and their respective charges) below input, with DGLAP splitting kernels at $\mathcal{O}(\alpha)$ in QED. Other such sets were subsequently developed that either adopted similar phenomenological models\cite{cteq}, or sought to constrain the photon in an analogous way to other partons by fits to Drell-Yan data \cite{nnpdf1,xfitter}, first developed by the NNPDF Collaboration. These early sets saw relatively large discrepancies between photon PDFs. 
Large modelling uncertainties persisted due to the freedom in the choice of scale above which photons are produced radiatively, modelled in the MRST set as the difference between the current and constituent quark masses, while the approach taken by the CTEQ14QED set \cite{cteq} was to attempt to fit a parameterisation based on the total momentum carried by the photon from $ep\rightarrow e\gamma + X$ data. In the case of NNPDF2.3QED~\cite{nnpdf1},
 the constraints available directly from data were rather weak, due to the small size of the PI contributions. This lead to large photon PDF errors, with a $\mathcal{O}(100\%)$ uncertainty at high $x$. In all cases the available data was unable to constrain the photon to a high degree of accuracy.

A final significant drawback of these early sets was that the majority did not account for the contribution to the photon PDF from elastic scattering, in which the proton coherently emits electromagnetic radiation without disintegration, in contrast to photon contributions previously accounted for from inelastic scattering processes, assumed to arise from quark splittings. This distinction between the elastic and inelastic photon emission was one that was seldom systematically treated, if considered at all. 

Significant strides have been made in recent years to overcome these deficiencies. First, more accurate determinations of the photon distribution at input have been developed by making use of the experimentally well determined elastic form factors of the proton, as in~\cite{martin_ryskin} and further developed in~\cite{Harland-Lang:2016apc,lucian}. 
More precisely, the photon PDF corresponds to the flux of emitted photons within the context of the equivalent photon approximation, and as discussed in some of the early work on this~\cite{EPA}, the contributions from elastic and inelastic emission to the photon PDF are directly related to the corresponding structure functions ($F_{1,2}^{ el}$, $F_{1,2}^{inel}$) probed in lepton--proton scattering. This idea has been revived in various works over the previous decades~\cite{Anlauf:1991wr,Blumlein:1993ef,Mukherjee:2003yh,Luszczak:2015aoa}, and has most recently been demonstrated within a rigorous and precise theoretical framework by the LUXqed group~ \cite{a,b}, where the first publicly available photon PDF applying this approach was also provided. 
As the elastic and inelastic proton structure functions have been determined experimentally to high precision, this has in turn allowed for the determination of the elastic and inelastic contributions to the photon to the level of a few percent. 
In addition to these developments, QED DGLAP splitting kernels have now been calculated to $\mathcal{O}(\alpha\alpha_S)$ \cite{alphaalphas} and $\mathcal{O}(\alpha^2)$ \cite{alpha2}, whose effects, as shown in Section~\ref{ssec_DGLAP}, are not insignificant to the evolution of the photon and other partons. 
In light of this, a greater confidence may be had regarding the effects of QED modified partons and their impact on cross section calculations.

In this paper, we outline the efforts undertaken by the MMHT group to develop a fully consistent set of QED partons, adopting the LUXqed formulation at input scale $Q_0$ for the photon. QED splitting kernels to $\mathcal{O}(\alpha)$, $\mathcal{O}(\alpha_S)$ and $\mathcal{O}(\alpha^2)$ are incorporated into the DGLAP evolution and the effect of this is explored. Furthermore, we also adopt a model for higher--twist (HT) effects in the quarks at low $Q^2$, as the evolution of the photon PDF is sensitive to these corrections, due to a lower input scale used in comparison to that of other PDF sets. 

As well as the conventional set of QED altered PDFs, we provide grids for the photon PDF separated into its elastic and inelastic components, as well as a consistent set of QED corrected neutron PDFs. Although the phenomenological implications of a neutron set are limited, their production is necessary for a consistent fit to deuteron and nuclear fixed target data from neutrino ($\nu$N) DIS scattering experiments used to constrain the PDFs. The QED corrected neutron PDFs of MRST \cite{mrstqed} provided isospin violating partons, with $u_{(p)} \neq d_{(n)}$, and these were seen to reduce the NuTeV $\sin^2\theta_W$ anomaly \cite{nutev_sin2}. The breaking of isospin symmetry may also have implications for the development of nuclear PDFs, and our current treatment develops this earlier approach, providing new predictions for the magnitude of isospin violation.

Finally, we will explore the phenomenological consequences of this set, demonstrating the effects of QED incorporation on $F_2(x,Q^2)$ as calculated from PDFs, the partonic luminosities as a function of centre-of-mass (CoM) energy and the change in fit quality after refitting the partons with QED. We also explore the consequences of fitting to ATLAS high-mass Drell-Yan data~\cite{DY-ATLAS-2}, with both QED effects and PI corrections to the cross section produced by our set. We find that the effect of a fully coupled QED DGLAP evolution is non-negligible on the gluon and quark PDFs.


\section{Including QED Effects in the MMHT Framework}\label{sec_qed}

In this section we describe how the MMHT framework has been modified to incorporate the QED splitting kernels in DGLAP evolution and the form we take for the input distribution of the photon, and discuss their effect on the final set of partons and the corresponding PDF uncertainties.

\subsection{Baseline QCD Fit}\label{ssec_qcd_basis}

Throughout this paper, in order to meaningfully interpret the effects of including QED effects, we will compare the new partons to a baseline set of PDFs evolved and fit solely with QCD kernels (at, unless explicitly stated, NNLO). However, this set differs from the most recent public release of partons, MMHT2014 \cite{MMHT}. In particular, this more closely corresponds to the set described in~\cite{MMHT2}, where the HERA Run I + II combined cross section data~\cite{HERA_I_II} have been included in the fit. Furthermore, we now include some additional data on $t\bar{t}$ production ($\sigma(t\bar{t})$) from the ATLAS and CMS collaborations. 
In addition, further small amendments have been made to the NLO and NNLO QCD kernels in the evolution, as detailed in Section~\ref{ssec_DGLAP}. Hence, we refer to this as the MMHT2015 PDF set and the PDFs with the QED effects included as MMHT2015qed.  

\subsection{Input Photon Distribution}\label{ssec_input}

To generate PDFs from QED corrected DGLAP evolution requires an input distribution for the photon at some starting scale, $Q_0$, from which the PDFs may be evolved to higher scales.
In principle, the photon input may be parameterised in a form similar to other partons, which in MMHT primarily uses an expansion in a basis of Chebyshev polynomials (as discussed in Section 2.1 of \cite{MMHT} and initially investigated in \cite{chebyshev}). Photon input distributions based on such an approach have been disfavoured by most groups due to the insufficient constraints provided directly from data when simultaneously fitting all of the partons. In particular, freely parameterising the photon (in a suitable expansion basis, analogous to the other partons) in a global fit is seen to lead to large uncertainties \cite{nnpdf1}.

As discussed above, a significantly more precise approach is to formulate the photon PDF in terms proton structure functions. This allow a precisely constrained input PDF  to be directly obtained from data for lepton-proton scattering; i.e. from the experimentally determined values of $F_2$ and $F_L$. 
We are always considering photon exchange in what follows, we will implicitly be referring to the Neutral Current (NC) structure functions wherever mentioned (i.e. $F_2 \equiv F_2^{NC}$, $F_L \equiv F_L^{NC}$). Moreover, as $Q^2_0 = 1\, \textrm{GeV}^2 \ll M_Z^2$ we can safely neglect any contributions from the weak neutral current and related interference terms.
The input expression for the photon PDF used in MMHT2015qed is derived from that of LUXqed~\cite{a} with some modification. At a given input scale, $\mu^2 = Q_0^2$, we take the photon PDF to be: 
 \begin{equation}\label{eq_input_lux}
        \begin{split}
            x\gamma(x,Q_0^2) = \frac{1}{2\pi\alpha(Q_0^2)}\int_x^1\frac{dz}{z}\Big\{ \int_{\frac{x^2 m_p^2}{1-z}}^{\frac{Q_0^2}{1-z}}\frac{dQ^2}{Q^2}\alpha^2(Q^2)\bigg[\bigg(zP_{\gamma,q}(z)+\frac{2x^2m_p^2}{Q^2}\bigg)F_2(x/z,Q^2)\\-z^2 F_L(x/z,Q^2)\bigg]-\alpha^2(Q_0^2)z^2F_2(x/z,Q_0^2)\Big\}, 
        \end{split}
\end{equation} where $\alpha=\alpha_{\rm QED}$ and $P_{\gamma,q}(z)$ corresponds to the $\mathcal{O}(\alpha)$ DGLAP splitting kernel given by:
    \begin{equation}
        P_{\gamma,q}(z) = \frac{1+(1-z)^2}{z}.
    \end{equation}
Note that the upper limit of the $Q^2$ integral
introduces a dependency on terms at scales higher than the input scale. It is more convenient to recast Eq. \eqref{eq_input_lux} such that the photon at input is purely dependent on contributions from $Q^2 < Q_0^2$, with all $Q^2 > Q_0^2$ dependence driven by DGLAP evolution. To achieve this, we separate the $Q^2$ range of the integral into two, with
 \begin{equation}\label{eq_input_lux2}
        \begin{split}
            x\gamma(x,Q_0^2) = \frac{1}{2\pi\alpha(Q_0^2)}\int_x^1\frac{dz}{z}\Big\{ \int_{\frac{x^2 m_p^2}{1-z}}^{Q_0^2}\frac{dQ^2}{Q^2}\alpha^2(Q^2)\bigg[\bigg(zP_{\gamma,q}(z)+\frac{2x^2m_p^2}{Q^2}\bigg)F_2(x/z,Q^2)\\-z^2 F_L(x/z,Q^2)\bigg]+\int_{Q_0^2}^{\frac{Q_0^2}{1-z}}\frac{dQ^2}{Q^2}\alpha^2(Q^2)\bigg[\bigg(zP_{\gamma,q}(z)+\frac{2x^2m_p^2}{Q^2}\bigg)F_2(x/z,Q^2)\bigg]\\-\alpha^2(Q_0^2)z^2F_2(x/z,Q_0^2)\Big\},
        \end{split}
\end{equation} 
where we have dropped the $F_L$ term in the second $Q^2$ integrand for simplicity. This can be justified on the grounds that $F_L \ll F_2$ and also by consideration of the fact that $F_L \sim \mathcal{O}(\alpha_S)$ in the parton model, while the expression given in Eq. \eqref{eq_input_lux} is formally only accurate to $\mathcal{O}(\alpha\alpha_S, \alpha^2)$. A more thorough discussion of this is given in Section 3 of \cite{b}.

\begin{figure}
\scalebox{.65}{\includegraphics{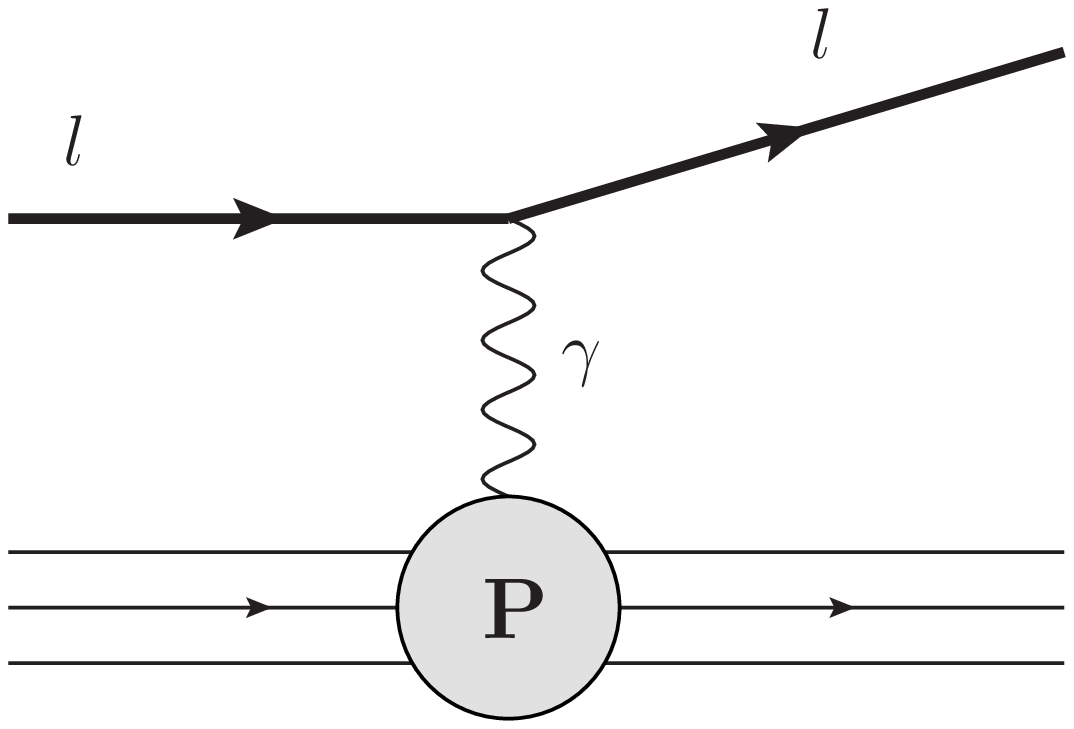}}
\begin{raggedright}\hspace{15mm}
\scalebox{.65}{\includegraphics{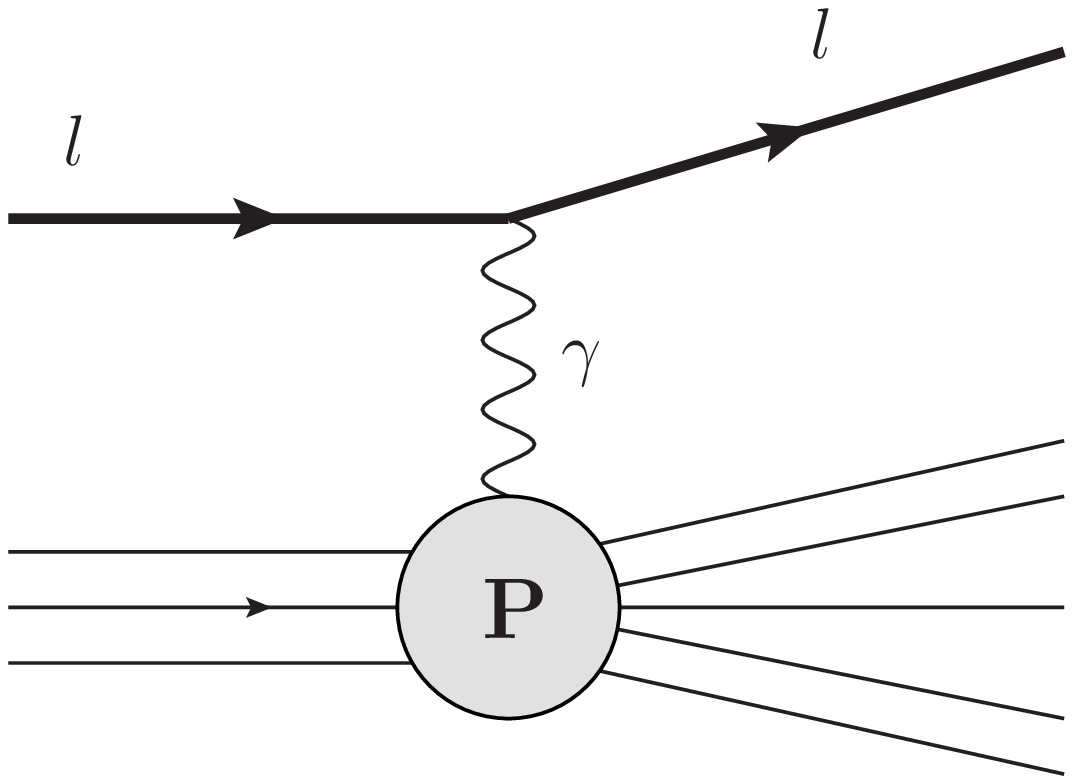}}
\end{raggedright}
\caption{Leading order representations of elastic (left) and inelastic (right) NC lepton--proton scattering processes.} 
\label{fig:DIS}
\end{figure}

By taking note of the fact that the scale variation of $F_2(Q^2)$ and $\alpha(Q^2)$ may be treated as stationary at the order we are calculating at ($\partial F_2/\partial Q^2$, $\partial\alpha/\partial Q^2 \sim 0$), we get  
 \begin{equation}\label{eq_input}
        \begin{split}
            x\gamma(x,Q_0^2) = \frac{1}{2\pi\alpha(Q_0^2)}\int_x^1\frac{dz}{z}\Big\{ \int_{\frac{x^2 m_p^2}{1-z}}^{Q_0^2}\frac{dQ^2}{Q^2}\alpha^2(Q^2)\bigg[\bigg(zP_{\gamma,q}(z)+\frac{2x^2m_p^2}{Q^2}\bigg)F_2(x/z,Q^2)\\-z^2 F_L(x/z,Q^2)\bigg]-\alpha^2(Q_0^2)\bigg(z^2+\ln(1-z) zP_{\gamma,q}(z)-\frac{2x^2m_p^2z}{Q_0^2}\bigg)F_2(x/z,Q_0^2)\Big\}.
        \end{split}
    \end{equation} 
This is the final expression for the input photon PDF that we will use throughout this paper, taking $Q_0^2=1\,{\rm GeV}^2$ as the input scale.
We note that this closely resembles 
 Eq.(4.10) of \cite{b}, 
 however in our case we retain the term of order $\mathcal{O}(m_p^2/Q_0^2)$ as this is more significant for the lower input scale we consider in comparison to LUXqed, which uses $Q_0^2 = 10\,{\rm GeV}^2$. 
We now elaborate on the composition of $F_{2,L}$ and how each source contributes to our expression for $x\gamma(x,Q_0^2)$. As discussed in the previous section, $F_{2,L}$ receive contributions from both elastic and inelastic scattering processes, 
as shown in Fig. \ref{fig:DIS}. In other words: \begin{equation}\label{eq_F2_split}
    F_{2,L} = F_{2,L}^{( el)} + F_{2,L}^{( inel)}.
\end{equation} 
As we will discuss below, the elastic and inelastic components of $F_{2,L}$ are obtained from fits to data, largely in the same way as in LUXqed~\cite{a,b}.

For $F_{2,L}^{(el)}$ we use the A1 collaboration fit \cite{c} to elastic scattering data, which is provided in terms of the Sachs electric and magnetic form factors for the proton:\begin{equation}\label{eq_sachs}
\begin{split}
    F_2^{( el)}(x,Q^2) = \frac{[G_E(Q^2)]^2 + \tau [G_M(Q^2)]^2}{1+\tau}\delta\Big(1-x\Big),\\
    F_L^{( el)}(x,Q^2) = \frac{[G_E(Q^2)]^2}{\tau}\delta\Big(1-x\Big),
\end{split}
\end{equation} 
where $\tau = Q^2/(4m_p^2)$. We note that the fits from the A1 collaboration differ from the widely used dipole approximation by about $10\%$ at $x\sim 0.5$; above this the difference increases further but this has little impact due to the effective kinematic cut at high $x$, discussed below.
However, as discussed in \cite{a}, the dipole model's reasonably good ($\mathcal{O}(5\%)$) correspondence to the data at low $x$ makes it useful in interpreting the scaling behaviour in this region ($\gamma^{(el)}(x) \sim \alpha \ln(1/x)$).

By substituting Eq. \eqref{eq_sachs} into Eq. \eqref{eq_input} we obtain an explicit formula for the elastic contribution to the photon PDF at a scale $\mu$,
\begin{equation}\label{eq_photel}
\begin{split}
    x\gamma^{(el)}(x,\mu^2) = \frac{1}{2\pi\alpha(\mu^2)x}\int_{\frac{x^2 m_p^2}{1-x}}^{\mu^2}\frac{dQ^2}{Q^2}\alpha^2(Q^2)\bigg[\bigg(xP_{\gamma,q}(x)+\frac{2x^2m_p^2}{Q^2}\bigg)\times\\\frac{[G_E(Q^2)]^2 + \tau [G_M(Q^2)]^2}{1+\tau}-x^2 \frac{[G_E(Q^2)]^2}{\tau}\bigg],
\end{split}
\end{equation}
which, noting the presence of the $1/\alpha(\mu^2)$ factor outside the integral, is equivalent to the order to which we calculate to solving the coupled DGLAP evolution for $\gamma^{el}$.

Turning to $F_{2, L}^{( inel)}$, 
this displays two distinct modes of behaviour. For the continuum $W^2 \gtrsim 4$ GeV\textsuperscript{2} region, the $x, Q^2$ dependence of $F_{2,L}$ is seen to be relatively smooth,
while in the resonance $W^2 \lesssim 3$ GeV\textsuperscript{2} region, various Breit-Wigner type resonances contribute, due to the presence of hadronic excited states such as the $\Delta$ and associated modes. To describe both of these regions, two different fits are used above and below a threshold of $W_{\rm cut}^2 = 3.5$ GeV\textsuperscript{2}.
For the continuum ($W^2 \geq W_{\rm cut}^2$) region, we use the HERMES GD11-P \cite{d} fit, while for the resonance ($W^2 < W_{\rm cut}^2$) region we take a fit to data from the CLAS collaboration \cite{e}. 

The HERMES collaboration \cite{d} provides data for $F_L$ by relating it to the available data for $F_2$. In particular, by considering the parameter $R = \sigma_L/\sigma_T$, the ratio of the longitudinal and transverse polarisation cross sections, the two structure functions are related in the following manner:\begin{equation}\label{eq_FL}
    F_L(x,Q^2) = \Big(1+\frac{4 m_p^2x^2}{Q^2}\Big)\frac{R(x,Q^2)}{1+R(x,Q^2)}F_2(x,Q^2)\;,
\end{equation}
where the function $R(x,Q^2)$, following the approach taken by HERMES, is adapted from the E143 collaboration fit, $R_{1999}$ \cite{R}. Although only $F_2$ data is provided by the CLAS fit, $F_L$ is estimated in the resonance region by using Eq. \eqref{eq_FL}, with the same form of $R(x,Q^2)$ provided by HERMES.

The structure functions themselves exhibit enhanced sensitivity to particular effects at lower starting scales (1 GeV in the MMHT framework, in comparison to 10 GeV adopted by LUXqed) such as proton mass corrections $\mathcal{O}(m_p^2/Q^2)$ and higher twist terms. Hence, modifications are made to account for these during the evolution, as discussed in subsection \ref{ssec_DGLAP}.

Finally, we note that the lower bound of the $Q^2$ integral in Eq.~\eqref{eq_input} introduces a cut on all photon contributions above a certain point in $x$. In particular, by noting that the integral in $z$ is bounded by $x$, at the limits of the integral the following inequality is imposed: 
\begin{equation}\label{eq_q2_cut}
    Q^2 \geq \frac{x^2m_p^2}{1-x},
\end{equation} 
which may be rearranged to express an upper limit on $x$ for $x\gamma(x,Q^2)$: 
\begin{equation}\label{eq_xcut}
    x \leq \frac{-Q^2 + Q\sqrt{Q^2+4m_p^2}}{2m_p^2} \equiv x_{\rm cut},
\end{equation} 
such that all contributions for $x > x_{\rm cut}$ vanish. As the expression at input, Eq. \eqref{eq_input} is valid at all scales, we include this cut at all stages of the evolution, not just for the input photon. As discussed in Section~\ref{ssec_DGLAP}, this leads to a dampening. 
As $Q\rightarrow\infty$, $x_{cut}\rightarrow 1$ and this constraint disappears rapidly, e.g. at $Q^2 = 10$ GeV\textsuperscript{2} we have $x_{\rm cut} = 0.918$. 
On the other hand, at the starting scale, we have $x_{ cut} \simeq 0.62$.
This cut has the effect of dampening the effects of other terms relevant to the photon 
evolution at high $x$, such as higher twist and target mass corrections, which are most 
prevalent at low $Q^2$, as well introducing a source of momentum sum rule violation, as discussed in Section~\ref{ssec_mom}.

\subsection{Modifications to DGLAP evolution}\label{ssec_DGLAP}

\subsubsection{PDF Basis}

In this section we outline the changes made to our evolution procedure to accommodate the effects of QED. First, we distinguish between the basis (the linearly independent combinations of partons) as parameterised in the fit and in the evolution. The partons are parameterised solely at the input scale, the majority of which, as discussed in subsection 2.1 of \cite{MMHT} and studied in \cite{chebyshev}, are based on an expansion in terms of Chebyshev polynomials ($T_i^{Ch}(y)$):\begin{equation}\label{eq_parameter}
    xf(x,Q_0^2) = A(1-x)^\eta x^\delta \Bigg(1+\sum\limits_{i=1}^{n}a_iT_i^{Ch}(y(x))\Bigg), 
\end{equation} with $y = 1-2\sqrt{x}$ and $n = 4$. Hence, the free parameters 
are $A, \eta, \delta$ and $a_i$, where some of the $A$ are fixed by sum rules.  
Distributions of this form are used for $f = u_{V}, d_{V}, S, (s+\bar{s)}$, where $S$ denotes the light-quark sea distribution:\begin{equation}\label{eq_sea}
    S = 2(\bar{u}+\bar{d})+s+\bar{s},
\end{equation} 
For the differences $\bar{d}-\bar{u}$ and $s-\bar{s}$ a reduced parameterisation is taken, reflecting the inability of current data to constrain these distributions to a high degree of precision. 
The gluon is provided in a form similar to Eq. \eqref{eq_parameter}, but with an additional term:\begin{equation}\label{eq_gluon}
    xg(x,Q_0^2)= A_g(1-x)^{\eta_g}x^{\delta_g}\Bigg(\sum_{n=1}^{2}a_{g,i}T_i^{Ch}(y(x))\Bigg) + A_{g'}(1-x)^{\eta_{g'}}x^{\delta_{g'}}\;,
\end{equation}
which is found to significantly improve the quality of the global fit~\cite{g_low_x}, and essentially provides more freedom for the gluon at low $x$. Here, $\eta_g$, $\eta_g'$, $A_g$ and $A_g'$ are all correlated in the fit, and their dependency can be artificially disrupted with the introduction of QED effects, leading to significant changes in the gluon if the partons are not refit, see Section \ref{ssec_qed_fit}.

In the MMHT framework the QED supplemented evolution of the partons is unidirectional in $Q^2$ from a starting scale of $Q_0 = 1$ GeV, with the convolution for the partons at each step performed in $x$ space. This is in contrast to that of NNPDF3.1luxQED case~\cite{nnpdf2}, which adopts an iterative process in its fit that reverses the DGLAP evolution of the partons from $Q = 100$ GeV for a photon produced from the NNPDF partons at high scales (with an elastic and low $Q^2$ contribution whose expression is the same as for LUXqed) then aims to find a consistent starting scale photon, subject to the momentum sum rule constraint for the partons, modified to include the photon ($\gamma$): 
\begin{equation}\label{eq_mom}
    \int_0^1 x(\Sigma(x,Q_0^2) + g(x,Q_0^2) + \gamma(x,Q_0^2)) = 1, 
\end{equation} 
where $\Sigma$ is the total singlet for the quarks. In practice, as we will see the resulting photon distributions from either approach are in agreement, differing only on the order of the uncertainties. However, due to certain higher twist effects and the procedure adopted for the treatment of our elastic photon distribution, $\gamma^{(el)}$, Eq. \eqref{eq_mom} is not strictly obeyed during the evolution, see Section \ref{ssec_mom} for further discussion.

While \crefrange{eq_parameter}{eq_gluon} reflect the input distribution parameterisations, a different and distinct linear combination of the partons are involved in the evolution procedure itself. Previously in the MMHT framework, the pure QCD DGLAP evolution of the partons, at all orders, was performed in a basis that was chosen for computational efficiency. This involved a decoupling of the partons into a singlet (consisting of the gluon and flavour combinations of quark and antiquark distributions) and non-singlet distributions which are evolved separately. Explicitly, the linearly independent combinations of partons that were evolved consisted of the following singlet (in the space of quark flavours) combinations: 
\begin{equation}
\label{SING}
\Sigma_L = u + \bar{u} + d + \bar{d} + s + \bar{s},
\end{equation}
\begin{equation}
\label{CPLUS_BPLUS}
c + \bar{c}, \hspace{20mm} b + \bar{b},
\end{equation}
\begin{equation}
\label{GLUE}
g
\end{equation} and the following non-singlet combinations:
\begin{equation}
\label{V_SUM}
u_{V} + d_{V} = u - \bar{u} + d - \bar{d},
\end{equation}
\begin{equation}
\label{NS_1}
u + \bar{u} - \frac{\Sigma_L}{3}, \hspace{20mm} -(s + \bar{s}) + \frac{\Sigma_L}{3},
\end{equation}
\begin{equation}
\label{NS_2}
\frac{d_{V}-u_{V}}{2}, \hspace{10mm}(s - \bar{s}) - \frac{u_{V}+d_{V}}{2}, \hspace{10mm} (c - \bar{c}) - \frac{u_{V}+d_{V}}{2},  \hspace{10mm}(b - \bar{b}) - \frac{u_{V}+d_{V}}{2},
\end{equation}
where the subscript $L$ in $\Sigma_L$ denotes the fact that the singlet consists only of the light quarks. The charm and bottom singlet distributions in Eq. \eqref{CPLUS_BPLUS} are evolved separately since they only become non-zero near the relevant mass thresholds for production.
 When considering QCD in isolation, the $SU(n_f)$ flavour invariance of the splitting kernels allows such distributions to be evolved consistently.

Now, the introduction of QED splitting kernels, $P_{ij}^{(QED)}$, in DGLAP evolution necessarily prohibits such combinations from being used. 
Writing the pure QCD splitting kernels via the usual perturbative expansion 
\begin{equation}
\label{qcd_splitting}
P_{ij}^{(QCD)} = \frac{\alpha_{S}}{2\pi}P_{ij}^{(1)} + \left(\frac{\alpha_{S}}{2\pi}\right)^{2}P_{ij}^{(2)} + \left(\frac{\alpha_{S}}{2\pi}\right)^{3}P_{ij}^{(3)} + ...
\end{equation}
for the QED and mixed QED--QCD case, recent theoretical work \cite{alphaalphas,alpha2} enable the terms
\begin{equation}
\label{qed_splitting}
P_{ij}^{(QED)} = \frac{\alpha}{2\pi}P_{ij}^{(0,1)} + \frac{\alpha\alpha_{S}}{(2\pi)^2}P_{ij}^{(1,1)} + \left(\frac{\alpha}{2\pi}\right)^{2}P_{ij}^{(0,2)} + ...
\end{equation}
to be used in the QED supplemented evolution. Here, the first and second superscript indices denote the order in QCD and QED respectively, and the second term in this expansion reflects mixed order splitting kernels.

Since the non-abelian nature of QCD does not manifest at leading order in quark interactions, the majority of the splitting functions in QCD and QED are simply related at this order:
\begin{align}\label{eq_QED_splittings}
P_{qq}^{(0,1)} &= \frac{e_{q}^2}{C_{F}}P_{qq}^{(1,0)}, &P_{q\gamma}^{(0,1)} &= \frac{e_{q}^2}{T_{F}}P_{qg}^{(1,0)} \;,\\ \label{eq_QED_splittings_2}
P_{\gamma q}^{(0,1)} &= \frac{e_{q}^2}{C_{F}}P_{gq}^{(1,0)},  &P_{\gamma \gamma} &= -\frac{2}{3}\sum_{i}^{n_F}e_{i}^2\delta(1-y),
\end{align}
The exception is $P_{\gamma\gamma}$, which differs considerably from the expression for $P_{gg}$, due to the purely gluonic contribution in the latter case.

A further caveat regarding $P_{\gamma\gamma}$ is that we only include quark loops, and not those due to leptons.  
 In the latter case, consistency would require the corresponding introduction of lepton PDFs, which in principle enter amongst the partons discussed so far, due to splittings of the form $\gamma \rightarrow l\bar{l}$. 
More precisely, for $Q^2 > m_{l}^2$, lepton splittings should also be incorporated into $P_{\gamma\gamma}$, such that the sum over quarks is modified to include the leptons:\begin{equation} \label{eq_leptsplit}
    \sum_i e_i^2 = N_C \sum_q^{n_F}e_q^2 + \sum_l^{n_L}e_l^2.
\end{equation} 
In our framework, we neglect the latter term which accounts for leptonic contributions to $P_{\gamma\gamma}$, since the contribution of the photon itself enters as an $\mathcal{O}(\alpha)$ correction to the PDFs, with the lepton contributions at $\mathcal{O}(\alpha^2)$, implying they are extremely suppressed. This was studied more extensively in \cite{lepton_pdfs} where it was found that the magnitude of the lepton distributions were many orders of magnitude below those of $x\gamma(x,Q^2)$, with negligible effects on the PDFs at the scales considered in this paper. 

However, we note that the LUXqed PDF set \cite{b} does include this contribution in the DGLAP evolution used to develop their $x\gamma(x,Q^2)$. Since the right hand side of Eq. \ref{eq_QED_splittings_2} is a $\delta(1-x)$ term multiplied by a negative coefficient, the extra contributions from the lepton splitting terms in DGLAP are anticipated to slightly reduce the magnitude of a photon whose evolution accounts for them (as one anticipates from the process $\gamma\rightarrow l\bar{l}$). 

Upon inspection of \crefrange{eq_QED_splittings}{eq_QED_splittings_2}, even at leading order it becomes apparent that the distributions in \crefrange{SING}{NS_2} cannot be used since QED couplings no longer support flavour symmetry, due to the charge separation of up and down type quarks ($e_u \neq e_d$). Furthermore, one anticipates based on this observation the breaking of isospin symmetry when comparing the valence distributions of the proton and neutron, as discussed in Section~\ref{ssec_neut_DGLAP}. To accommodate the requirement of charge sensitivity, the partons are now evolved in the following basis, which are separable by charge: 
\begin{equation}
\label{eq_qplusminus}
q^{\pm}_{i} = q_{i} \pm \bar{q_{i}},\hspace{5mm} g,\hspace{5mm}\gamma^{(el)},\hspace{5mm}\gamma^{(inel)}.
\end{equation}
In the following discussion the subscript $i$ denotes any active ($Q>2m_q$) flavour: $i = u, d, s, c, b$ and the +/- superscript denotes the singlet and non-singlet quark distributions respectively. The gluon and photon components, $g$, $\gamma^{(el)}$ and $\gamma^{(inel)}$ are then evolved individually in the flavour space of the partons. 

Although the basis given in Eq. \eqref{eq_qplusminus} is compatible with a joint evolution in QCD and QED, they require some modification to the form of DGLAP splitting kernels used. Writing $t = \ln(Q^2)$ and $(f\otimes g)(x) = \int_{x}^{1} \frac{dy}{y}f (x/y) g(y)$, the non-singlet distributions described in Eqs. \eqref{V_SUM} - \eqref{NS_2} may be evolved in the following way in pure QCD:
\begin{equation}
\label{eq_NS_evo}
\frac{\partial q_{i}^{NS}}{\partial t} = P_{q_{i}}^{-}\otimes q_{i}^{NS}\;,
\end{equation}
where the expression for $P_{q_{i}}^{-}$ may be found in Eqs. (4.94) to (4.108) of \cite{pinkbook}. The simplicity of this equation arises from the fact that symmetry allows for evolution of the $p_i^{NS}$ distributions to be diagonal in quark flavour space, such that only the term $P_{q_{i}}^{-}$, which describes the diagonal elements of the quark-quark and quark-antiquark splitting functions, is required. 

The evolution for the $q_{i}^{-}$ requires an additional component since although they are also non-singlet functions of the quarks, the non-diagonal elements in flavour space become necessary to the evolution:
\begin{equation}
\label{eq_qminus_evo}
\frac{\partial q_{i}^{-}}{\partial t} = P_{q_{i}}^{-}\otimes q_{i}^{-} + \sum_{j=1}^{n_{F}}\Delta P^{S}\otimes q_{j}^{-},
\end{equation}
where $\Delta P^S$ becomes non-zero at NNLO ($\mathcal{O}(\alpha^3)$) in QCD and $n_F$ is the number of active quarks in the evolution.

Note that this sum over valence-like non-singlet distributions corresponded to eq. \eqref{V_SUM} in the original MMHT framework, which neglected the strange, charm and bottom distributions due to their small relative size. With the release of the set described in this paper, the contribution from these off-diagonal splittings for all flavours are now included, which represent minor changes, $\mathcal{O}(10^{-5})$, in a like-for-like comparison with the original MMHT partons purely in QCD.

\subsubsection{Target Mass and Higher Twist Corrections}

As previously noted, MMHT2015qed differs in its production of a photon PDF from other contemporary sets in adopting a straightforward evolution in $Q^2$ space, from a starting scale of $Q_0 = 1$ GeV. However, at low scales such as these, target mass corrections, which account for the finite mass of the proton, and higher twist terms have non-negligible contributions to $F_{2,L}$. Above $Q_0$, the $F_2^{(inel)}$ contributions to $\gamma^{(inel)}$, as in eq. \eqref{eq_input}, are modelled by the parton splittings in DGLAP, which require some modification to capture the relevant behaviour at high $x$.

The target mass corrections for the proton are well known, modifying the $\mathcal{O}(\alpha)$ quark to photon splitting in an identical manner to the first term in the integrand of Eq. \eqref{eq_input}:
\begin{equation}\label{eq_target_mass}
    P_{\gamma,q}^{(0,1)}(z) \rightarrow P_{\gamma,q}^{(0,1)}(z) + \frac{2x^2m_p^2}{zQ^2}.
\end{equation}
Further modifications are also required for higher twist terms which lead to discrepancies between $F_2$ as calculated from the partons and experimental measurements for $F_2^{(inel)}$, due to non-perturbative effects at high $x$ and low $Q^2$. 
In a global fit this effect is typically eliminated by cutting on the low $W^2$ region where such corrections are relevant, however for the determination of the photon PDF 
which is sensitive to $F_2^{({ inel})}$ in the region discussed, we must include this. 
Therefore a phenomenological model must be adopted to account for such higher--twist corrections.
We follow the approach of~\cite{i}, where non-perturbative $\sim 1/Q^2$ power corrections to the structure functions are provided, by characterising the associated infrared divergences in field theory with the so--called renormalon. In this paper we shall use the term renormalon synonymously with higher twist corrections of this type. In \cite{i}, they provide at $\mathcal{O}(1/Q^2)$ a modification to $F_2$ that accounts for the change due to renormalon calculations at high $x$, and this is found to give an improved description of DIS data \cite{renorm_data}.

In lieu of $F_2^{({ inel})}$, during the evolution the contributions to $\gamma^{({ inel})}$ are essentially generated by the quark splittings ($q\rightarrow q \gamma$), where the total quark singlet $\Sigma$ 
plays the role of $F_2$ in eq. \eqref{eq_input}. Therefore, to approximate renormalon effects during the evolution, these modification are instead made to the quarks via
\begin{equation}\label{eq_renormalon}
q(x,Q^2) \rightarrow q(x,Q^2)\Big(1 + \frac{A'_2}{Q^2}\int^1_{x}\frac{dz}{z}C_2(z)q(\frac{x}{z},Q^2)\Big) ,
\end{equation} 
where $A'_2$ is a parameter not given a priori by the theory and $C_2(z)$ is defined in Eq. 4.1 of \cite{i}, and conserves the flavour number properties of the various $q(x/z,Q^2)$. As such, higher twist contributions to $F_2$ do not contribute to the Adler sum rule, 
\begin{equation}
\int_0^1{ d}x\,F^{HT}_2(x,Q^2) = 0\;,
\end{equation}
enforcing that these are well behaved as $x\rightarrow0$. However no such restriction applies to $F_3$, and renormalon calculations \cite{f3_HT} imply that they become large, necessitating the need for the more stringent cut on $F_3$ data used in the fit (from the CHORUS collaboration \cite{chorus}) that extend into this region. 
\begin{figure}
\centerline{%
\scalebox{0.7}{\input{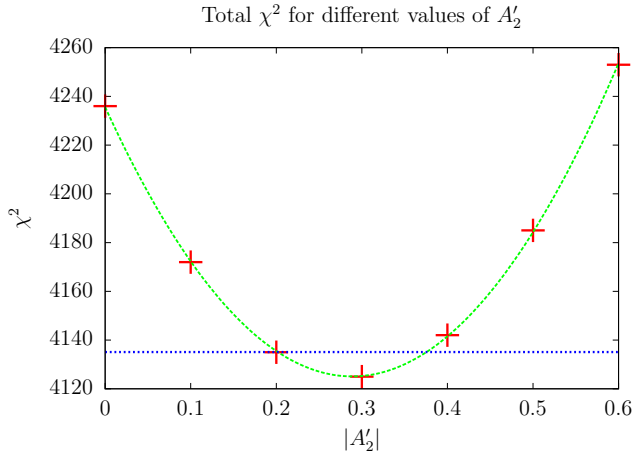}}}
\caption{The $\chi^2$ values obtained in a global fit, with kinematic constraints on DIS data lowered to $W^2 = 5$ GeV\textsuperscript{2}, with different values of $A'_2$ in the renormalon calculations for $F_{2}$ and $F_{3}$ . The dashed blue line represents a $\Delta \chi^2$ = 10 variation from the minimum to establish an uncertainty band on $A'_2$.}
\label{fig_A_2}
\end{figure}
This is of interest because the parameter $A'_2$ is not well determined, and in \cite{i}, is fit loosely to structure function data to yield a value of $A'_2 = -0.2$ GeV\textsuperscript{2}.
As discussed above, data sensitive to renormalon contributions are typically excluded in global fits, to remove any sensitivity to such non--perturbative effects. 
In particular, in MMHT kinematic cuts of $W^2 > 15$ GeV\textsuperscript{2} (and $W^2 > 20$ GeV\textsuperscript{2} at LO) are taken, while for those data sets relating to $\nu(\bar{\nu})N$ experiments to measure $xF_3$ a more stringent cut of $W^2 > 25$ GeV is imposed~\cite{MSTW}.
However, with the aim of determining a more precise value of $A'_2$ we have relaxed these constraints, lowering the threshold to $W^2 > 5$ GeV\textsuperscript{2} and modifying $F_{2}$ and $F_{3}$ to include the relevant renormalon contributions as in \cite{i}, i.e. with modifications of the form shown in eq. \eqref{eq_renormalon}. 

In Fig. \ref{fig_A_2} the fit quality for different values of $A'_2$ is shown. We find that $A'_2 = -(0.3\pm 0.1)$ GeV\textsuperscript{2}, with uncertainties determined from a generous $\Delta \chi^2 = \pm 10$ variation in the fit (to one significant figure). This is motivated by the dynamical tolerance scheme used in our framework, as outlined in Section 6 of \cite{MSTW}, where it was found that in order to provide reasonable uncertainties when fitting to many disparate data sets in tension with one another, one typically requires tolerances $T = \sqrt{\Delta\chi^2_{global}} \sim 3$
 rather than the $T = 1$ one would obtain from a standard `parameter-fitting' criterion. 
We note this choice also corresponds to the fixed tolerance uncertainty schemes adopted by early CTEQ sets~\cite{cteq_tolerance}. The uncertainty on this is then propagated as an independent source of uncertainty for the photon, as discussed in Section~\ref{ssec_unc}. This represents a slightly larger renormalon contribution than predicted from \cite{i}, though the data are unable to provide significant constraints in either case.

\begin{figure}
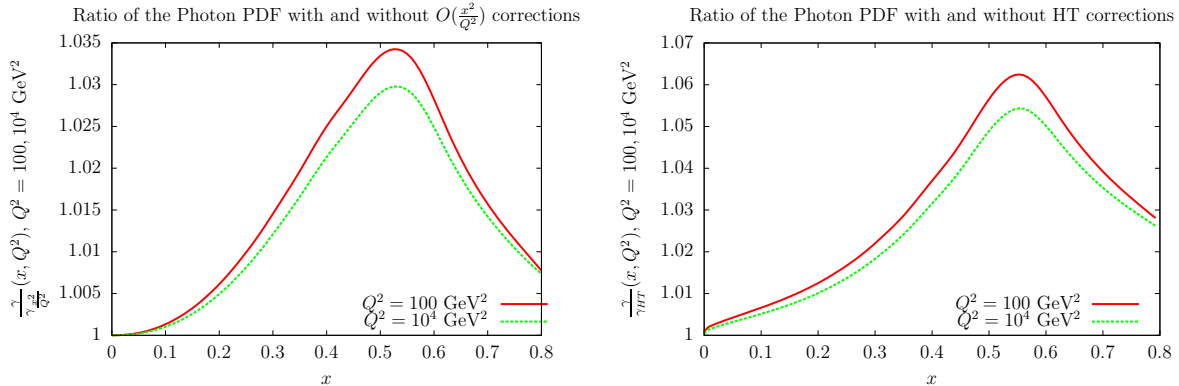

\centering
\begin{minipage}{.5\textwidth}  
  \centering
  \scalebox{0.6}{\input{PHOT_MP_RAT.tex}}
\end{minipage}%
\begin{minipage}{.5\textwidth}
  \centering
  \scalebox{0.6}{\input{PHOT_REN_RAT.tex}}
\end{minipage}
\caption{(Left) Ratio of the photon PDF with ($\gamma(x,Q^2)$) and without ($\gamma_{\frac{x^2}{Q^2}}(x,Q^2)$) target mass corrections and (right) Higher Twist (renormalon) corrections.}
\label{fig:mass}
\end{figure}

As seen in Fig. \ref{fig:mass}, the target mass corrections lead to a $\sim 3\%$ increase in the photon at high $x$, while the renormalon contributions, which provide an increasingly positive contribution to $F_2$ at high $x$, correspondingly enhance the photon at moderate to high $x$. Note that the turn around in both figures at $x \simeq 0.5$ occurs due to the previously mentioned effective kinematic cut on all photon contributions at high $x$ and low $Q^2$.
This cut itself is also a function of the proton mass $m_p$, though for our purposes we consider the kinematic cut imposed due to the target mass (i.e. the cut in $x$) as independent from the term introduced in the evolution and it is seen that the two have opposite effects on the high $x$ photon, with the kinematic cut ultimately dominating and the effect of the corrections to the splitting function and the renormalon contribution being suppressed as $x \to 1$.

Since the target mass and renormalon contributions are both $\sim 1/Q^2$ corrections, their relative importance at higher scales is seen to decrease slightly, as shown by a comparison of the red ($Q^2 = 100$ GeV\textsuperscript{2}) and green ($Q^2 = 10^4$ GeV\textsuperscript{2}) curves. We note that both the proton mass term and the modification to the quarks in eq. \eqref{eq_renormalon} introduce small, independent sources of momentum violation in the evolution, as discussed in Section~\ref{ssec_mom}. 

\subsection{Separation of Elastic and Inelastic Components}\label{ssec_split}

\begin{figure}
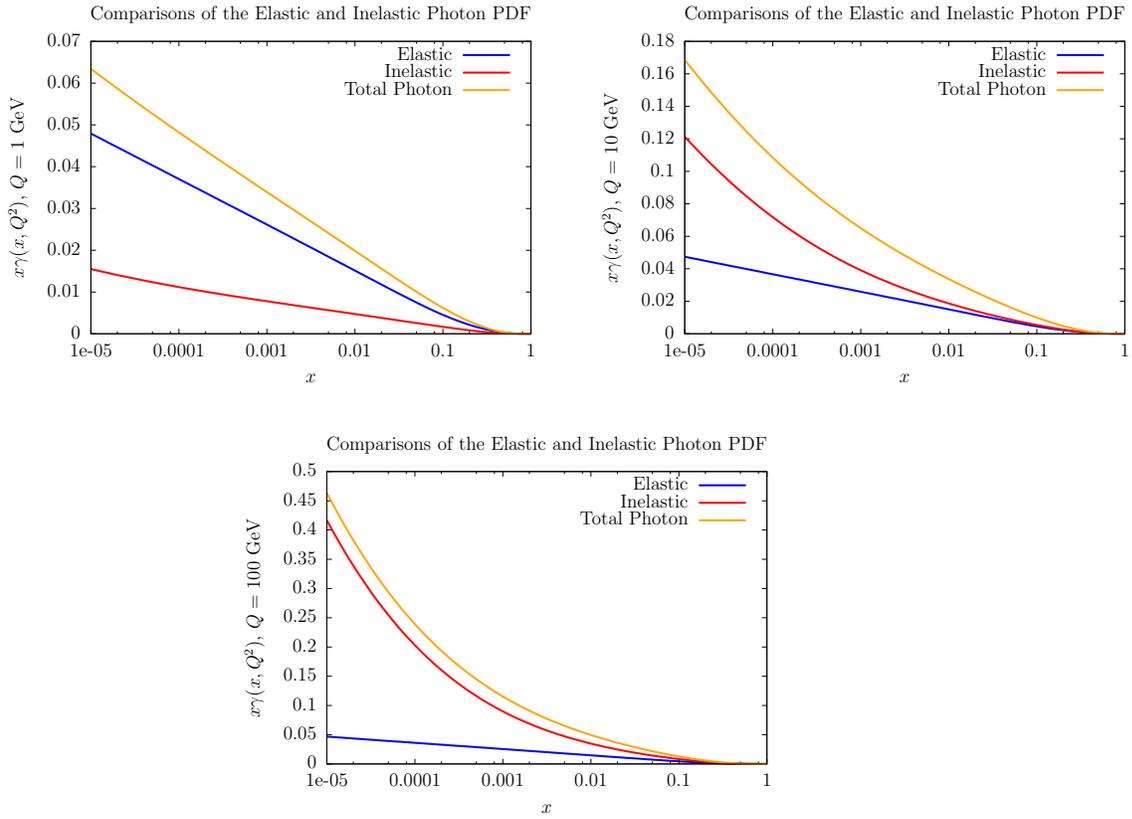

\scalebox{.6}{\input{phot_1.tex}}
\begin{raggedright}
\scalebox{.6}{\input{phot_10.tex}}
\end{raggedright}
\begin{flushleft}
\hspace{3cm}
\scalebox{.6}{\input{phot_100.tex}}
\end{flushleft}
\caption{The Elastic and Inelastic Photon components at different values of $Q$. Top left: $Q = 1$ GeV, top right: $Q = 10$ GeV, bottom: $Q = 100$ GeV.}
\label{fig:elastic}
\end{figure} 

As noted in Section~\ref{ssec_input}, the photon PDF actually comprises of two component distributions, $\gamma (x,Q^2) = \gamma^{(el)}(x,Q^2) + \gamma^{(inel)}(x,Q^2)$, which represent photon contributions from elastic and inelastic proton scattering events, respectively. Separating $\gamma^{(el)}$ and $\gamma^{(inel)}$ from one another while consistently performing the evolution for all the partons required certain changes to be made from the standard procedure for performing DGLAP, due to the fact that the generation of $\gamma^{(el)}$ in the evolution is independent of parton splittings, as detailed below.

For $\gamma^{(inel)}$, the evolution is analogous to that of the other partons. The contributions from the HERMES (continuum) and CLAS (resonance) data for $F_2^{(inel)}$ are present only at input, above which DGLAP evolution is performed. We emphasise that all photon contributions that arise from the splitting of other partons (the quarks, antiquarks and both photon components themselves, but also the gluon at $\mathcal{O}(\alpha\alpha_S)$) in DGLAP are absorbed into the definition of $\gamma^{(inel)}$ (using the notation of the previous section):\begin{equation}
    \frac{d\gamma^{(inel)}}{dt} = \sum_j^{n_F}P_{\gamma q_j}\otimes q_j + \sum_j^{n_F}P_{\gamma\bar{q}_j}\otimes \bar{q}_j + P_{\gamma g}\otimes g + P_{\gamma\gamma}\otimes\gamma^{(inel)}.
\end{equation} This reflects the fact that scattering processes that are sensitive to the partons are themselves inelastic and that therefore any photon contributions that arise from their evolution in DGLAP are necessarily inelastic contributions.

While $\gamma^{(el)}$ is included at input and passed to the other partons during evolution, its own evolution requires consideration of the contributions it receives above $Q_0$ from $F_2^{(el)}$, since our expression for $\gamma^{(el)}$ given in Section~\ref{ssec_input}, Eq. \eqref{eq_photel}, holds generally above the input scale. Incorporating this and splittings of the form $\gamma\rightarrow q \bar{q}$ and $\gamma\rightarrow q \bar{q} g$ at $\mathcal{O}(\alpha\alpha_S)$, the evolution for $\gamma^{(el)}$ is given as:
\begin{equation}\label{eq_photel_evol}
    \frac{d\gamma^{(el)}}{dt} = P_{\gamma\gamma}\otimes\gamma^{(el)} +  \delta x\gamma^{(el)}.
\end{equation} 
The expression for $ \delta x\gamma^{(el)}$ is given by taking the derivative of the expression for the elastic photon, eq. \eqref{eq_photel}, w.r.t $Q^2$:
\begin{equation}\label{eq_photcoh}
\begin{split}
        \delta x\gamma^{(el)}(x,Q^2) = \frac{\alpha(Q^2)}{2\pi}\frac{1}{x}\bigg[\bigg(xP_{\gamma,q}(x)+\frac{2x^2m_p^2}{Q^2}\bigg)\frac{[G_E(Q^2)]^2 + \tau [G_M(Q^2)]^2}{1+\tau}\\-x^2 \frac{[G_E(Q^2)]^2}{\tau}\bigg].
\end{split}
\end{equation}
As discussed in the next section, including the term introduced in Eq. \eqref{eq_photcoh} as an external contribution (not generated from parton splittings but added into the evolution from $F_2^{(el)}$ data) introduces a small amount of momentum violation, as do subsequent splittings of the form $\gamma^{(el)}\rightarrow q\bar{q}$.

Although the provisions outlined above are needed for the evolutions of $\gamma^{(el)}$ and $\gamma^{(inel)}$, i.e. those contributions from splitting functions of the form $P_{\gamma\{q,\bar{q},g,\gamma\}}$, the treatment for the rest of the partons remains broadly unchanged.
Since the quark, antiquark and gluon contributions from $P_{\{q,\bar{q},g,\gamma\}\gamma}$ splittings do not distinguish between $\gamma^{(el)}$ and $\gamma^{(inel)}$, the entire photon contribution, $\gamma(x,Q^2) = \gamma^{(el)}(x,Q^2) + \gamma^{(inel)}(x,Q^2)$, is passed to the relevant splitting kernels during evolution. 

\begin{figure}
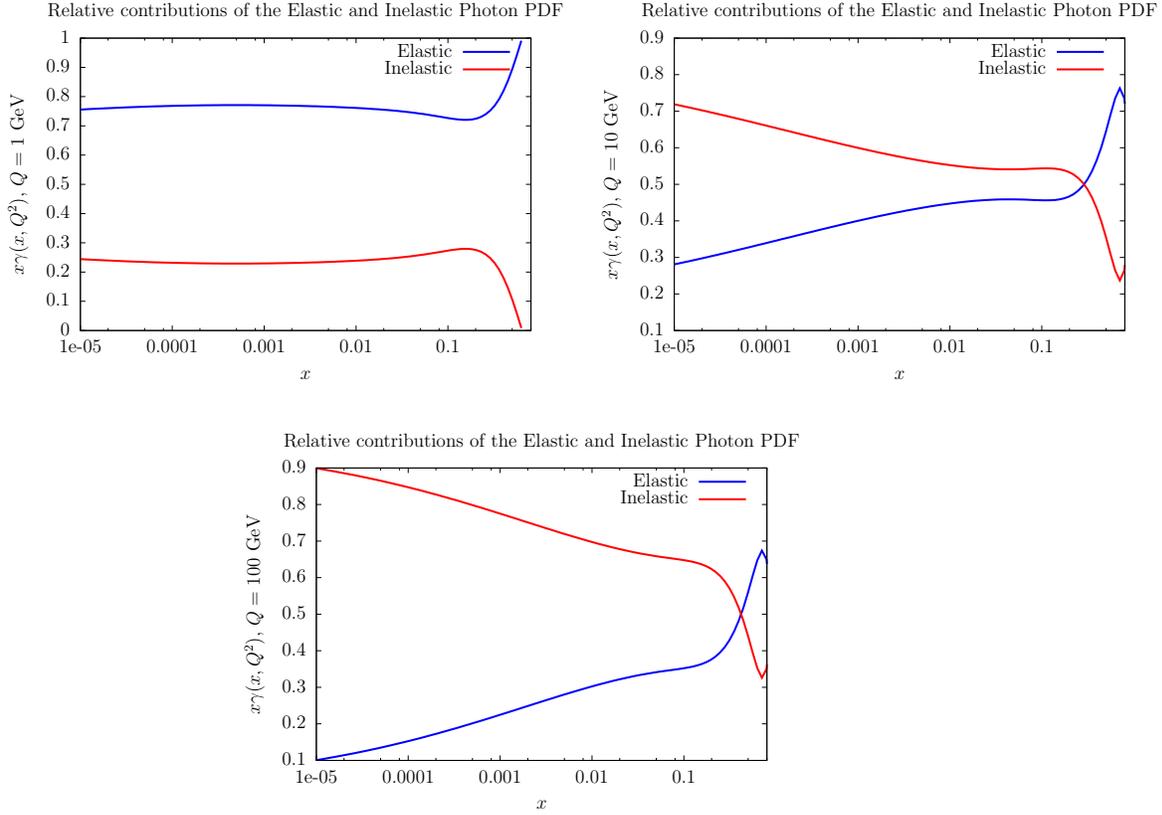

\scalebox{.6}{\input{phot_1_rat.tex}}
\begin{raggedright}
\scalebox{.6}{\input{phot_10_rat.tex}}
\end{raggedright}
\begin{flushleft}
\hspace{3cm}
\scalebox{.6}{\input{phot_100_rat.tex}}
\end{flushleft}
\caption{The relative contributions of the Elastic and Inelastic Photon components at different values of $Q$. Top left: $Q = 1$ GeV, top right: $Q = 10$ GeV, bottom: $Q = 100$ GeV.}
\label{fig:elastic_rat}
\end{figure}

\begin{figure}
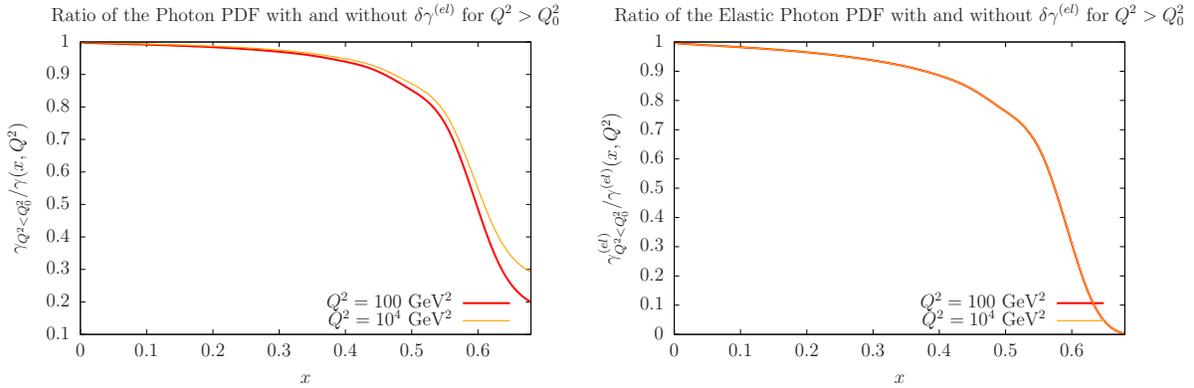

\scalebox{.6}{\input{phot_nocoh_rat.tex}}
\begin{raggedright}
\scalebox{.6}{\input{photcoh_nocoh_rat.tex}}
\end{raggedright}
\caption{A plot showing the ratio of the photon distribution with the 
elastic contribution for $Q^2> Q_0^2$ removed and the total distribution.  
The right plot shows the same ratio for only the elastic contribution 
to the photon distribution.}
\label{fig:coh_above}
\end{figure}

As $\gamma^{(el)}$ and $\gamma^{(inel)}$ distinguish between the photon in two distinct categories of scattering processes, there is a phenomenological interest in comparing the two. At input, the elastic contribution dominates over that of the inelastic, as $F_2^{(el)} > F_2^{(inel)}$ in the region $Q \lesssim 1$ GeV. However, evolution quickly enhances the contributions of $\gamma^{(inel)}$, particularly at low $x$, predominantly due to quark splittings, as shown in Figs. \ref{fig:elastic} and \ref{fig:elastic_rat}.
As discussed above, the only contributions $\gamma^{(el)}$ receives during the evolution are those from Eq. \eqref{eq_photcoh}. Since $G_{E,M}(Q^2)$ are known to diminish with increasing $Q^2$ and $1/\tau \sim 1/Q^2$, an inspection of the form of eq. \eqref{eq_photcoh} reveals that it will be of diminishing importance in a significant range of $x$. In fact, investigating the effects of leaving out this term in eq. \eqref{eq_photel_evol} entirely yielded a $\gamma^{(el)}$ with differences of just $\mathcal{O}(10^{-3})$ from the form with the contributions included. 
However, the elastic distribution's contribution at input, and above, is proportionally large at high $x$, even at high $Q^2$ (Fig. \ref{fig:elastic_rat}), and due to the kinematic cut all contributions to the photon at the highest $x$ are from $Q^2>Q_0^2$. Indeed, in this region the elastic contribution even above $Q_0^2$ dominates the photon, as can be seen in Fig. \ref{fig:coh_above}, which shows the effect on the photon of turning this contribution off. One slight caveat, however, is that as $\lim_{x\to1} \gamma^{(el)}, \gamma^{(inel)} \rightarrow 0$, and ultimately uncertainties become large in this region (see Section~\ref{ssec_unc}), making it difficult to make very strong predictive statements about either distribution in this region.

\subsection{Momentum Conservation}\label{ssec_mom}

The inclusion of the photon PDF requires that the photon be included in the momentum sum rule~\eqref{eq_mom}, naturally leading to a redistribution of momentum in the other partons in order to obey eq. \eqref{eq_mom} at input. However, due to the procedure adopted for the inclusion of $\gamma^{(el)}$, outlined in the previous section, as well as higher twist terms, this equation is not strictly obeyed during the evolution. This reflects the discrepancy between effects of non-perturbative corrections, such as that of target masses, and the parton model. In this section we outline the consequences of such changes.

\begin{figure}
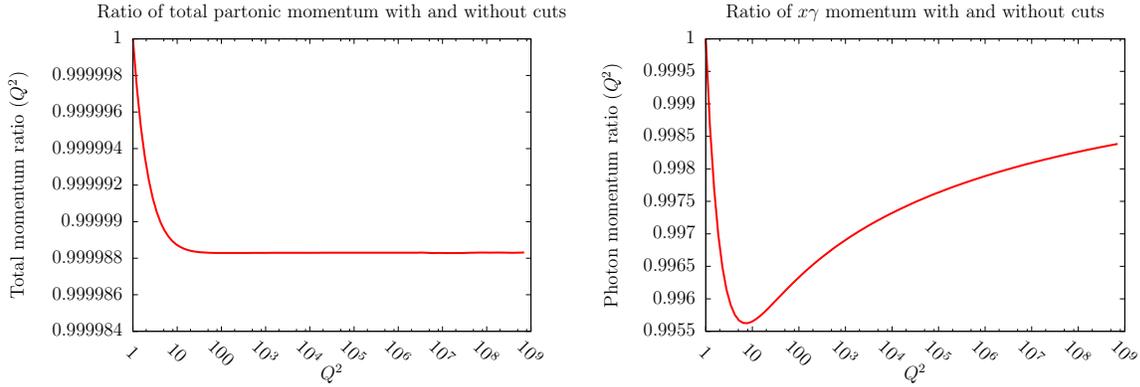

\scalebox{.6}{\input{momentum_rat_cut.tex}}
\begin{raggedright}
\scalebox{.6}{\input{momentum_rat_cut_phot.tex}}
\end{raggedright}
\caption{(Left) A plot showing the ratio between the total momentum carried by all the partons, at a given point in $Q^2$ of the evolution with the kinematic cut $Q^2 \geq x^2m_p^2 / (1-x)$ applied to $x\gamma(x,Q^2)$ and without. The right plot shows an identical plot but focusing solely on the proportional difference in momenta caused by the photon, where the effects on the evolution are seen to peak at $Q^2 \sim 10$ GeV\textsuperscript{2}.}
\label{fig:mom_cut}
\end{figure}

First we discuss the effect of a kinematic cut on the photon, as introduced by the lower limit of the integral in $Q^2$ in the expression for $x\gamma(x,Q^2)$, which as discussed in Section \ref{ssec_input} has the effect of introducing an effective cut on the photon PDF at high $x$ during the evolution. In essence, this removal of photon contributions at high $x$ is a target mass correction (since the cut has a dependence on $m_p^2$), which is not required to obey the momentum conservation of the partons ordinarily found in DGLAP evolution and therefore introduces small amounts of violation (in the form of a reduction of total momentum carried by the partons) into eq. \eqref{eq_mom} of 
$\mathcal{O}(10^{-3}\,\%)$. This is seen in Fig. \ref{fig:mom_cut} (left), where we display the ratio of the total momentum of the partons with and without this cut applied.

In particular, Fig. \ref{fig:mom_cut} (right) indicates that the reduction to the total momentum carried by the photon is, as anticipated, most strongly affected by the kinematic cut at low scales until $Q^2\sim 10$ GeV\textsuperscript{2} (with total changes of less than 1\%). Since the overall momentum carried by the photon is small, $\sim 2-3\times 10^{-3}$, at low scales where momentum violating effects are most prevalent, this leads to the minuscule amount of change observed in the total momentum of the partons.

We now discuss other effects during the evolution which contribute to violation of the momentum sum rule. 
The momentum sum rule is constrained to be obeyed by all the partons at the input scale, and both the inelastic and elastic photons are considered when imposing the momentum sum rule for the parameterisation of the quarks, as in eq. \eqref{eq_mom}.
However, above the input scale the contribution to $\gamma^{(el)}$ that comes from the second term in \eqref{eq_photel_evol}, that is due to elastic photon emission, will lead to some momentum sum rule violation, as this contribution does not originate from standard DGLAP evolution, and is not balanced by a corresponding loss of quark and antiquark momentum,  
i.e., any $\gamma$ contribution from the quarks during evolution, e.g. $q \rightarrow q+\gamma$ is absorbed into the definition of $\gamma^{(inel)}$. 
($\gamma^{(el)}$ is not entirely decoupled from the evolution of the quarks, since $\gamma^{(el)} \rightarrow q\Bar{q}$ splitting are still permissible.)
In practice, this effect is negligible, with momentum violating effects of $O(10^{-4})$ observed in the sum rule during evolution, and in fact stabilises at higher $Q^2$  where the elastic contribution is less significant. 

Similarly, the proton mass term given in eq. \eqref{eq_target_mass} naturally breaks the form of momentum conservation usually obeyed between splitting functions of this type, implied by the equation\begin{equation}\label{eq_split_mom_con}
    \int_0^1 x\Big[ P_{q,q}^{(0,1)}(x) + P_{\gamma,q}^{(0,1)}(x)\Big] = 0.
\end{equation}
In essence, the proton mass term invalidates this relationship, though in rapidly diminishing amounts as $1/Q^2\rightarrow 0$, leading to changes of $\mathcal{O}(10^{-5})$ in the total momentum carried by the partons.

\begin{figure}
\centerline{%
\scalebox{0.7}{\input{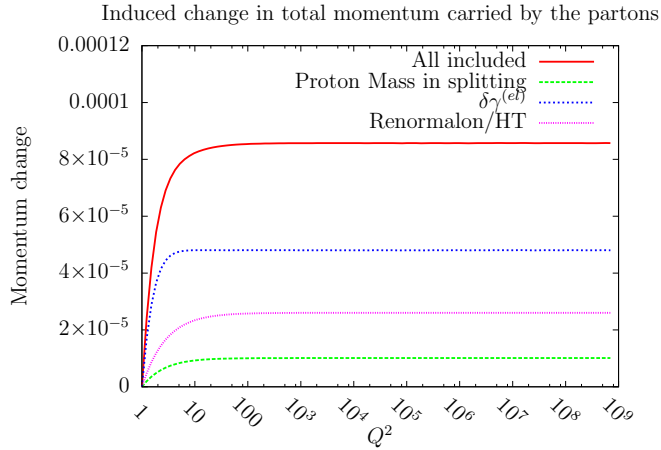}}}
\caption{The absolute change induced in the total momentum carried by the partons due individually to the target mass corrections, $\mathcal{O}(m_p^2/Q^2)$, the inclusion of $\delta\gamma^{(el)}$ in eq. \eqref{eq_photel_evol} (elastic contributions above input) and Higher Twist/Renormalon contributions to $\gamma^{(inel)}$. }
\label{fig:mom_con}
\end{figure}

Likewise, other higher twist terms included in the evolution for the purposes of QED lead to small amounts of momentum violation. Since the quark distributions, $q_i(x,Q^2)$, passed to both $P_{q,q}^{(0,1)}(x)$ and $P_{\gamma,q}^{(0,1)}(x)$ differ due to the inclusion of renormalon corrections for the latter but not the former, this aspect of the evolution also invalidates momentum violation to a small degree, also shown in Fig. \ref{fig:mom_con}, creating a small amount of violation of $\mathcal{O}(2\times10^{-5})$.

Overall, even in conjunction, the combined magnitude of momentum sum rule violation 
is less than $10^{-4}$. In practice, this total effect is less than the momentum violation coming from the `leakage' of the partons that occurs due to the fact that the integration range during DGLAP does not strictly begin at $0$ for the convolutions of $xf(x/z,Q^2)$ with the splitting functions, which are instead defined in the MMHT framework only to a finite level of precision (defined at a lower bound of $x \sim 10^{-12}$). Therefore, we do not consider any of the effects described above as serious invalidations of the parton model, even with the full spectrum of effects due to QED included. 

\section{QED Neutron PDFs}\label{sec_neutron}

Neutron PDFs are necessary for interpreting the results of deuterium scattering experiments that are still widely used in PDF fits to constrain the flavour decomposition of the proton.
The most widely adopted approach is to assume isospin symmetry between hadrons in the valence distributions:
\begin{equation}
\label{eq_isospin}
u_{V,(p)}(x,Q^2) = u_{(p)}(x,Q^2) - \bar{u}_{(p)}(x,Q^2) = d_{V,(n)}(x,Q^2) = d_{(n)}(x,Q^2) - \bar{d}_{(n)}(x,Q^2),
\end{equation}
\begin{equation}
\label{eq_isospin2}
d_{V,(p)}(x,Q^2) = d_{(p)}(x,Q^2) - \bar{d}_{(p)}(x,Q^2) = u_{V,(n)}(x,Q^2) = u_{(n)}(x,Q^2) - \bar{u}_{(n)}(x,Q^2),
\end{equation} 
where the subscripts \{(p),(n)\} denote the proton and neutron respectively. 
In practice, this is seen to produce a good agreement with the observed data and is well motivated by the $SU(n_f)$ flavour symmetry of QCD as well as the fact that the evolution treats both quark flavours as essentially massless ($m_u^2/Q^2, m_d^2/Q^2 \sim 0$ for $Q > 1$ GeV).
 However, as discussed in Section~\ref{ssec_DGLAP}, QED splitting kernels such as those in eqs. \eqref{eq_QED_splittings},  \eqref{eq_QED_splittings_2} no longer uphold this symmetry and are expected to generate $\mathcal{O}(\alpha)$ violations in the above relations. Therefore, to relate the distributions of the proton to those of the neutron in a manner consistent with QED evolution, one needs to carefully account for the effects of the relevant quark charges $e_u, e_d$ in the evolution and to allow for small amounts of isospin violation to be introduced.

\subsection{Modified DGLAP Evolution}\label{ssec_neut_DGLAP}

QED corrections automatically result in isospin violating effects such that at given $x$ and $Q^2$ values, the valence distributions can no longer be related to one another by eqs. \eqref{eq_isospin}, \eqref{eq_isospin2}. However, any modification to these relations must still preserve the flavour quantum numbers of the proton and neutron via the usual sum rules
\begin{equation}\label{eq_qn_rule}
    \begin{split}
        \int_0^1 \mathrm{d}x \, u_{V,(p)}(x) = \int_0^1 \mathrm{d}x \,d_{V,(n)}(x)  = 2\\ \int_0^1 \mathrm{d}x \,  d_{V,(p)}(x)= \int_0^1 \mathrm{d}x \,  u_{V,(n)}(x)= 1.
    \end{split}
\end{equation}
Above input, one can in principle keep track of all contributions to the quarks that arise from QED splittings. In the case of the valence distributions, the evolution is governed by eq. \eqref{eq_NS_evo}, where the splitting kernels are separated into QED and QCD contributions via
\begin{equation}
P_{i,j} = P_{i,j}^{(QCD)} + P_{i,j}^{(QED)}\;.
\end{equation}
Therefore, one can distinguish between two contributions to the valence distributions in the proton (which we refer to as $q_V$ in the following discussion):
\begin{equation}\label{eq_valence_split}
    q_{V}(x,Q^2) = q^{(QCD)}_{V}(x,Q^2) + q^{(QED)}_{V}(x,Q^2),
\end{equation} 
where $q^{(QED)}_{V}$ is defined as:
\begin{equation}\label{eq_qed_valence}
    q^{(QED)}_{V}(x,Q^2) = \int_{Q_{0}^2}^{Q^2} \frac{d\mu^2}{\mu^2} \frac{\alpha(\mu^2)}{2\pi}\Big(P_{q_{i}}^{-(QED)}\otimes q_{V}\big(\frac{x}{z},\mu^2\big)\Big), 
\end{equation}
and the integrand 
contains all QED splitting contributions to the valence distributions. 
Note that an implicit overall factor of the quark electric charge, $e_{q_i}^2$, is contained in $P_{q_{i}}^{-(QED)}$. 

To parameterise the isospin violating components 
 between the proton and the neutron, we define: 
\begin{equation}\label{eq_isospin_violating}
    \begin{split}
        \Delta d_{V,(n)}(x,Q^2) = d_{V,(n)}(x,Q^2) - u_{V,(p)}(x,Q^2)\\ \Delta u_{V,(n)}(x,Q^2) = u_{V,(n)}(x,Q^2) - d_{V,(p)}(x,Q^2), 
    \end{split}
\end{equation} 
where na{\"i}ve pointwise isospin conservation would lead both of these expressions to evaluate to $0$. 
For isospin violation generated by QED splittings, we assume that
\begin{equation}\label{eq_neutron_approx}
    \Delta d_{V,(n)}(x,Q^2) \propto u_{V,(p)}^{(QED)}(x,Q^2), \hspace{10mm} \Delta u_{V,(n)}(x,Q^2) \propto  d_{V,(p)}^{(QED)}(x,Q^2).
\end{equation} 
In particular, we assume that provided that the momentum and number conservation rules (eqs. \eqref{eq_mom}, \eqref{eq_qn_rule}) are obeyed by the constant of proportionality, the only further step needed in relating the valence distributions of the proton to that of the neutron is the charge re-weighting of the relevant valence distributions, $q_{V,(p)}^{(QED)}$, to correct for charge proportional terms in the evolution. Then, we may rewrite eq. \eqref{eq_isospin_violating} in the form of the following equations:
\begin{equation}\label{eq_delta_dv}
\Delta d_{V,(n)}(x,Q_0^2) = \epsilon\Big(1-\frac{e_{d}^2}{e_{u}^2}\Big)u_{V,(p)}^{(QED)}(x,Q_0^2),
\end{equation}
\begin{equation}\label{eq_delta_uv} 
\Delta u_{V,(n)}(x,Q_0^2) = \epsilon\Big(1-\frac{e_{u}^2}{e_{d}^2}\Big)d_{V,(p)}^{(QED)}(x,Q_0^2).
\end{equation} 
where $\epsilon$ is fixed to conserve momentum at input.

In order to satisfy momentum conservation, eq. \eqref{eq_mom}, at input for the neutron, one needs the neutron photon distribution at input.
This defines the constant of proportionality, $\epsilon$, by:
\begin{equation}\label{eq_epsilon}
\epsilon = \frac{\int_{0}^{1} dx x(\gamma_{(p)}(x)-\gamma_{(n)}(x))}{\int_{0}^{1} dx x(\frac{3}{4}u_{V,(p)}^{(QED)}(x) - 3 d_{V,(p)}^{(QED)}(x))}
\end{equation} 
where all the distributions are evaluated at $Q_0^2 = 1$ GeV\textsuperscript{2}. This follows a procedure similar to that adopted in \cite{mrstqed}. 

This expression implicitly depends on the assumption that the remaining partons are then related to one another in the standard manner, assuming that the antiquark (or sea) distributions are still well approximated by \begin{equation}\label{eq_neut_sing}
    (\bar{u})_{(n)}(x,Q_0^2) = (\bar{d})_{(p)}(x,Q_0^2), \hspace{10mm} (\bar{d})_{(n)}(x,Q_0^2) = (\bar{u})_{(p)}(x,Q_0^2)
\end{equation} 
with all other quark flavours and the gluon being related identically between hadrons.

Using eqs. \eqref{eq_delta_dv} and \eqref{eq_delta_uv} the $u$ and $d$ singlet distributions are then related to one another between hadrons by: 
\begin{equation}\label{eq_d_sing_n}
    (d+\bar{d})_{(n)}(x,Q^2) = (u+\bar{u})_{(p)}(x,Q^2)+\Delta d_{V,(n)}(x,Q^2)
\end{equation}
\begin{equation}\label{eq_u_sing_n}
    (u+\bar{u})_{(n)}(x,Q^2) = (d+\bar{d})_{(p)}(x,Q^2)+\Delta u_{V,(n)}(x,Q^2),
\end{equation} where $\Delta \{d,u\}_{V,(n)}$ are as defined above. Though of less apparent interest, these relations pertain to the discussion in Section~\ref{ssec_neut_photon}, where the neutron photon PDF is considered as primary sensitive to distributions of the type $q+\bar{q}$ during the evolution. In anticipation of this, we note that $\Delta \{d,u\}_{V,(n)}$ lead to differences between the isospin related $u$ and $d$ singlet distributions between hadrons of only $\mathcal{O}(1\%)$, since the $\Delta q_V$ terms are proportional to the contributions to the valence quarks that arise solely from QED evolution, which are $\mathcal{O}(\alpha)$ suppressed. In practice, relating these distributions to one another by isospin symmetry still remains a good approximation. This will underpin our development of a photon PDF of the neutron in the next section.

\begin{figure}
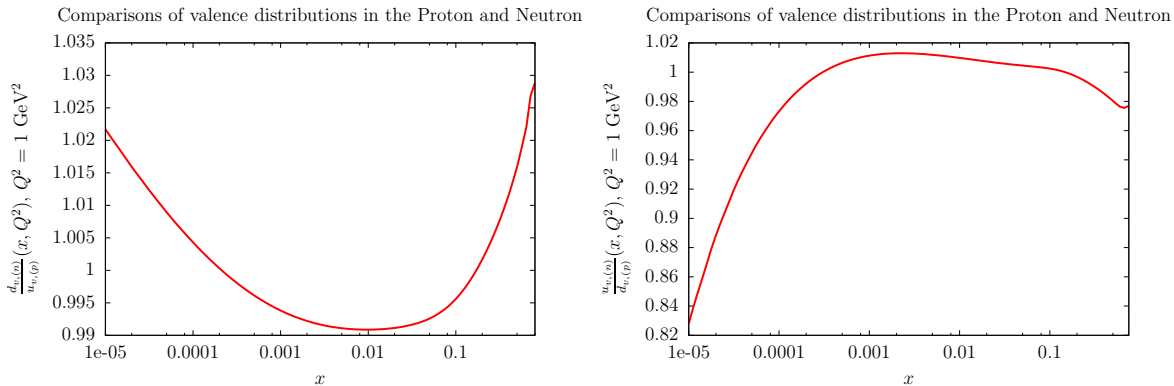

\scalebox{.6}{\input{UV_QED_1_RAT.tex}}
\begin{raggedright}
\scalebox{.6}{\input{DV_QED_1_RAT.tex}}
\end{raggedright}
\caption{The ratio of valence quarks, related to one another by isospin, of the neutron to that of the proton at the input scale $Q_0^2 = 1$ GeV\textsuperscript{2}. On the left is $u_{V,(n)}/d_{V,(p)}$, and on the right is $d_{V,(n)}/u_{V,(p)}$, both as functions of $x$.}
\label{fig:neutron}
\end{figure} 

For the valence distributions, in practice, the magnitude of isospin violation is seen to be a few percent, becoming significant especially at low and high $x$, where all distributions tend towards $0$, as shown in Fig. \ref{fig:neutron}. Of note is the fact that the discrepancy between the predicted ratio of valence quarks and the na\"{i}ve isospin assumption remains at the $\sim 1\%$ level, even for the peak of the valence distributions (at $x\sim\frac{1}{3}$, $x\sim\frac{2}{3})$). This effect is seen to increase during the evolution, with differences of $\sim5\%$ at $Q = 100$ GeV\textsuperscript{2}.   

Finally, although the primarily interest in this paper for the development of QED corrected neutron PDFs is to provide a manner of relating the PDFs to deuterium scattering experiments used to constrain the partons, we also wish to highlight the potential relevance of this set in the determination of nuclear PDFs. In particular, the assumption made in modern determinations of nuclear PDFs (such as those of EPPS \cite{nuclear_PDF_1} and nCTEQ \cite{nuclear_PDF_2}) is to fit to data with the assumption that the $u$ and $d$ quark type distributions in the neutron and proton are related to one another by isospin symmetry. With the development of this set, we propose that this assumption need not be applied strictly and that with the introduction of QED effects, the small amounts of isospin violation shown in Fig. \ref{fig:neutron} may be of relevance when the determination of nuclear PDFs reach the $\mathcal{O}(5\%)$ level. While current determinations do not reach this level of precision, a QED corrected relationship between proton and neutron PDFs may provide better fits to the available data, and is of interest given that recent work has begun to adopt quark flavour dependence in fits \cite{nuclear_PDF_3}.

\subsection{The Photon PDF of the Neutron}\label{ssec_neut_photon}

As in the case of the proton, there is also a corresponding photon PDF of the neutron, $\gamma_{(n)}(x,Q^2)$, which should in general be included.
At input, the expression for this is adapted from that of the proton, eq. \eqref{eq_input}, with the proton mass replaced by that of the neutron and the relevant form factors substituted or approximated in the manner discussed below.

As in the case of the proton, the input distribution is due to both inelastic and elastic photon emission.
\begin{figure}
\centerline{%
\scalebox{0.7}{\input{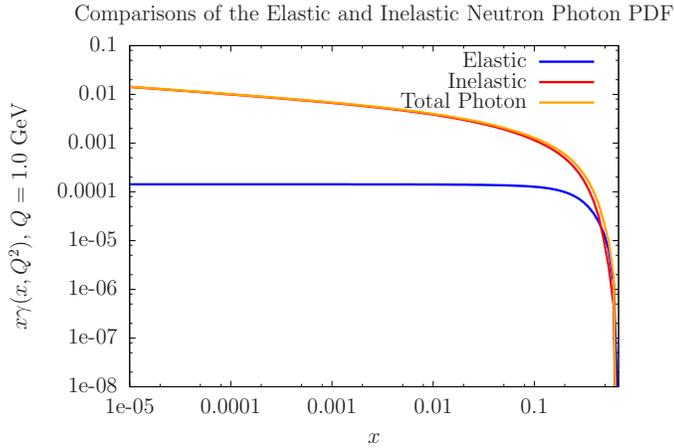}}}
\caption{The Elastic and Inelastic Photon components at $Q_0^2 = 1$ GeV\textsuperscript{2}.}
\label{fig_neut_input}
\end{figure}
The neutron elastic distribution $\gamma_{(n)}^{(el)}$ at input is given in terms of the Sachs form factors of the neutron, $G_{E,(n)}, G_{M,(n)}$.
We adopt the phenomenological Galster parameterisation~\cite{ge_n}:
\begin{equation}
    G_{E,(n)} = \frac{A\tau}{1+B\tau}G_D(Q^2),
\end{equation} 
where $\tau = Q^2/4m_n^2$ and $G_D(Q^2)$ is the dipole form factor for hadrons 
(in the form commonly used to approximate $G_{E,(p)}$ when multiplied by the proton's magnetic moment, $G_{E,(p)} = \mu_p G_D(Q^2)$):
\begin{equation}
    G_D(Q^2) = \frac{1}{(1+\frac{Q^2}{\Lambda^2})^2},
\end{equation}
 with $\Lambda^2 = 0.71$ GeV\textsuperscript{2}. Values for $A$ and $B$ are then taken from a fit to deuterium and $^3He$ scattering experiments 
  provided by \cite{ge_n_2}, for which
  \begin{equation}
  A = 1.70\pm 0.04, \quad B = 3.30\pm 0.32\;.
\end{equation}
For $G_{M,(n)}$ meanwhile, a simple dipole approximation of the form: 
\begin{equation}
    G_{M,(n)} = \mu_n  G_D(Q^2)\;,
\end{equation} 
is used. 
These are found~\cite{gm_n} to give a reasonably good 
fit to data provided from deuterium scattering experiments.

\begin{figure}
\centerline{%
\scalebox{0.7}{\input{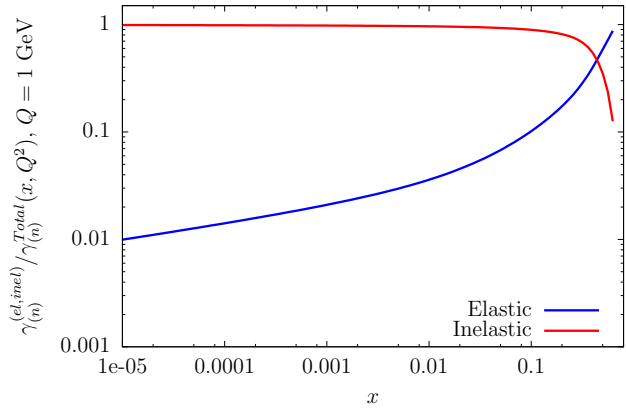}}}
\caption{The relative proportions of the Elastic and Inelastic Photon component contributions to the total Neutron Photon PDF at $Q_0^2 = 1$ GeV\textsuperscript{2}}
\label{fig:neut_input_rat}
\end{figure}

Due to the net neutral charge of the neutron, both form factors are significantly smaller in magnitude than those of the proton, and one therefore expects the relevant elastic contribution to $\gamma_{(n)}$ at input to be significantly smaller. In fact, as seen in Figs. \ref{fig_neut_input} and \ref{fig:neut_input_rat} it is found to scarcely contribute at all, comprising $\mathcal{O}(1\%)$ of the total photon over a large range of $x$, becoming significant only at $x\sim 0.5$, where the magnitude of the PDF itself is of vanishing importance. Therefore, given the uncertainties associated with both models adopted for both $G_{E,(p)}$ and $G_{M,(p)}$, $\gamma^{(el)}_{(n)}$ may reasonably be omitted for phenomenological purposes.
 This is even more true for contributions above input, since further elastic contributions are attenuated as $1/Q^2 \rightarrow 0$ such that $\gamma_{(n)}^{(el)}/\gamma_{(n)} \rightarrow 0$.

For the proton $F_{2}^{(inel)}$ can be divided into resonance and continuum contributions.
 In the resonance region, the fit provided by CLAS for $F_2^{(inel)}$ is also given for the neutron, and so can be straightforwardly applied here. 
For the continuum region however, 
 the HERMES fit for $F_2$ is provided solely for the proton. Therefore, for $F_{2,(n)}^{(inel)}$ we instead relate this approximately to the proton case.
 In particular, we re-weight the proton continuum contribution according to
 \begin{equation}
 F_{2,(n)}^{(inel)} = r_{F_2}\times F_{2,(p)}^{(inel)}\;,
 \end{equation}
 where $r_{F_2}$ is  the ratio of the charge weighted singlet of partons $\Sigma$, at input, for the neutron to that of the proton: 
\begin{equation}\label{eq_charge_weight}
r_{F_2} = \frac{4(d+\bar{d}) + (u+\bar{u}) + (s+\bar{s})}{4(u+\bar{u}) + (d+\bar{d}) + (s+\bar{s})}.
\end{equation}
It should be noted that in the expression above, all the distributions refer to those of the proton, where we have used the assumption (which as discussed in the previous section holds to a high degree of accuracy) that 
\begin{equation}
(u+\bar{u})_{(n)} = (d+\bar{d})_{(p)},  \quad (d+\bar{d})_{(n)} = (u+\bar{u})_{(p)}. 
\end{equation}
Note that any attempt to improve the accuracy of the expression in eq. \eqref{eq_charge_weight} by using eqs. \eqref{eq_d_sing_n}, \eqref{eq_u_sing_n} would not be feasible in the current framework since those equations depend on the parameter $\epsilon$ from the previous section, which in turn is determined from $\gamma_{(n)}^{(inel)}$ itself.  

\begin{figure}
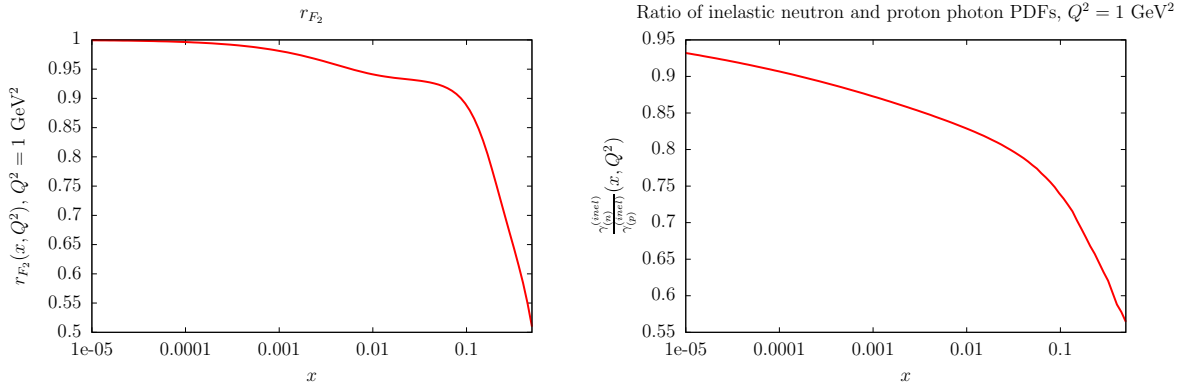

\scalebox{.6}{\input{rF2.tex}}
\begin{raggedright}
\scalebox{.6}{\input{INEL_INPUT_RAT.tex}}
\end{raggedright}
\caption{(Left) the ratio of the charge weighted light quark singlets between the neutron and proton. (Right) the ratio of $x\gamma^{(inel)}$ between the neutron and proton, for comparison.}
\label{fig:neutron_inel}
\end{figure}

By approximating the ratio of structure functions between the hadrons by their respective quark singlets, the form of $F_{2,(n)}^{(inel)}$ substituted in eq. \eqref{eq_input} for $\gamma_{(n)}^{(inel)}$ in the continuum region at input is simply given as $F_{2,(n)}^{(inel)} = r_{F_2}\times F_{2,(p)}^{(inel)}$. 
In Fig. \ref{fig:neutron_inel}, one sees a broad correspondence between $r_{F_2}$ and the ratio $\gamma^{(inel)}_{(n)}/\gamma^{(inel)}_{(p)}(x,Q_0^2)$, particularly as $x\rightarrow 0$, where the continuum region is dominant. Some discrepancy between the two plots exists due to the presence of the resonance region contribution, which as stated above is reformulated based on available neutron data, rather than being re-scaled by $r_{F_2}$. The general similarly however persists because at low $x$, the behaviour of the light quark singlets are dominated by the sea quarks, and $u \simeq \bar{u} \simeq d \simeq \bar{d}$, such that the effect of swapping flavours via isospin leaves the PDFs roughly invariant in this region.

Above input, $\gamma_{(n)}$ is approximated from the evolution of $\gamma_{(p)}$ in a manner analogous to that of the quarks as described in Section~\ref{ssec_neut_DGLAP}. 
We also distinguish between the flavour of the quark whose splitting leads to the evolution of the photon. 
One can label the contributions to $\gamma$ from the originating quark or antiquark flavour to obtain $\gamma_q$, given by the following expression\begin{equation}
         \gamma(x, \mu^2)_{q} = \int_{Q_{0}^2}^{\mu^2} \frac{\alpha(Q^2)}{2\pi}\frac{dQ^2}{Q^2}\int_{x}^{1}\frac{dz}{z}\Big(P_{\gamma,q}(z)q^+(\frac{x}{z},Q^2)\Big).
\end{equation} where the $+$ superscript once again denotes singlet type distributions of the form $q+\bar{q}$. 

Assuming isospin symmetry, which as shown in the previous section holds to a good approximation, one can make assumptions based on the predicted splittings in the neutron evolution to re-weight the contributions of each $\gamma_q$ of the proton, based on the scheme laid out in eq. \eqref{eq_neut_sing} to obtain:

\begin{equation}
    \begin{split}
    \gamma(x,Q^2)_{(n)}^{(inel)} =  \frac{e_d^2}{e_u^2}\gamma_{u,(p)}(x,Q^2) + \frac{e_u^2}{e_d^2}\gamma_{d,(p)}(x,Q^2) + \gamma_{\{s,c,b,g\},(p)}(x,Q^2)\;,
    \end{split}
\end{equation} where the final term accounts for all other flavours, whose contributions are assumed to be identical for the neutron and the proton. 

At the level of approximation adopted, the expression given above is expected to be accurate to $\mathcal{O}(\alpha)$, with errors of $\mathcal{O}(\alpha\alpha_S+\alpha^2)$. Anticipating results from the next section, it is seen that these higher orders induce changes in the resultant photon of $\sim 3\%$ at high $x$, while the uncertainties on the CLAS fit and the PDFs themselves each introduce a $\sim 1\%$ uncertainty on the photon PDF at low and high $x$ respectively. Therefore, one can conservatively estimate the uncertainty of the photon PDF of the neutron to be $\mathcal{O}(5\%)$ at high $x$ and $\mathcal{O}(2-3\%)$ at low $x$ where the PDF and higher order uncertainties dominate.

\begin{figure}
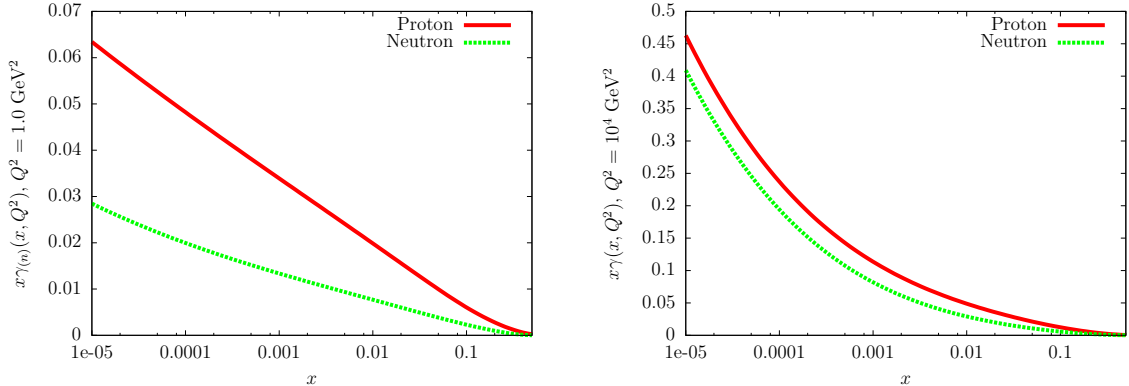

\scalebox{.6}{\input{PHOT_NEUT_VS_NEUT_1.tex}}
\begin{raggedright}
\scalebox{.6}{\input{PHOT_NEUT_VS_PROT_100.tex}}
\end{raggedright}
\caption{A comparison of the total Neutron and Photon PDFs at $Q = 1$ GeV (left) and $Q = 100$ GeV (right).}
\label{fig:neutrona}
\end{figure} 

As seen in Fig.~\ref{fig:neutrona}, at the input scale the photon PDF in the neutron is a factor of $\sim 2$ smaller than in the proton case, while for $Q = 100^2$ GeV\textsuperscript{2} the PDFs are comparable in size.
This is as expected, since the ratio of charges used to re-weight the proton contributions are $\mathcal{O}(1)$, and as $\gamma^{(el)}_{(p)}$ becomes less significant in the evolution, as seen in Fig. \ref{fig:elastic}, the inelastic contribution dominates.
\begin{figure}
\centerline{%
\scalebox{0.7}{\input{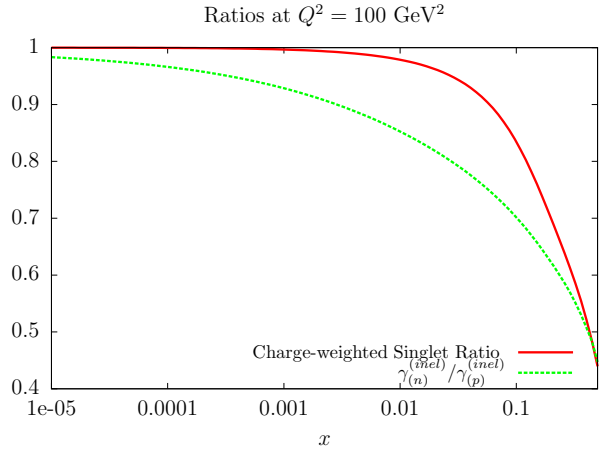}}}
\caption{The upper (red) curve is the ratio of $\sum_i e_{q_i}^2(q_i+\bar{q}_i)$, that is, the charge weighted sum of quark singlets, in the neutron to that of the proton. The lower (green) curve is the ratio of inelastic photon PDFs between the neutron and photon. At low $x$, these are both seen to tend towards unity as the flavour invariance of the sea (which obeys isospin symmetry maximally) dominates.}
\label{fig:neut_inel_100}
\end{figure}
This is seen in Fig. \ref{fig:neut_inel_100},
which shows the ratio of the charged-weighted quark singlets ($\Sigma_C$) between the proton and neutron, and the ratio of $\gamma^{(inel)}_{(n)}/\gamma^{(inel)}_{(p)}(x,Q^2)$ at the same scale. As shown above for the input, the isospin invariance demonstrated at low $x$ in the sea quarks means that the valence properties of the hadrons are less relevant at higher scales, leading to a photon PDF of the neutron that is comparable to that of the proton.

\section{Results}\label{sec_fit}

We now discuss the effect of adding QED corrections to the global PDF analysis.
First, in subsection 4.1, we present the changes to the PDFs due to including the QED corrections into the input and the evolution and we show the proton PDFs obtained from the new global analysis. In subsection 4.2.1 we discuss the global fit quality and in 4.2.2 we present the photon PDF and compare with other 
contemporary analyses. Also, in subsection 4.2.3 we show the QED corrected structure functions. In subsection 4.2.4 We briefly outline the impact of QED corrections on the best-fit value of $\alpha_S$. Subsection 4.2.5 then finishes with a presentation of the photon-photon luminosities
in $pp$ collisions. In subsection 4.3 we quantify the uncertainties in our determination of the photon PDF of the proton.

\subsection{Changes to PDFs due to QED corrections}\label{ssec_qed_fit}

Here we show the changes in the parton distributions that are produced as a result of the changes given in the preceding sections. We include the $\mathcal{O}(\alpha)$, $\mathcal{O}(\alpha\alpha_S)$ and $\mathcal{O}(\alpha^2)$ QED corrections, unless otherwise stated, and compare against the baseline PDF set without QED effects described in Section~\ref{ssec_qcd_basis}.

\begin{figure}[h]
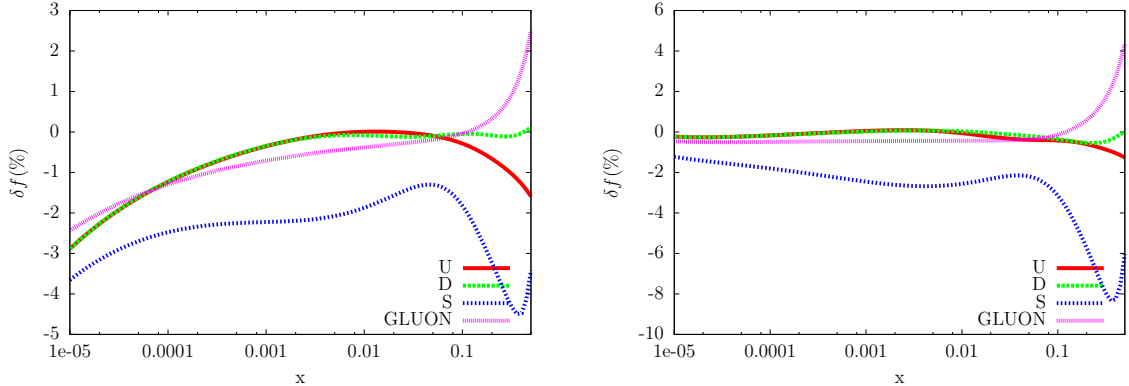

\scalebox{.6}{\input{PARTON_QED.tex}}
\begin{raggedright}
\scalebox{.6}{\input{PARTON_QED_FIT.tex}}
\end{raggedright}
\caption{The percentage change in the $u, d, s, g$ partons at $Q = 100$ GeV due to QED evolution with (right) and without (left) refitting to data.}
\label{fig:parton_mom}
\end{figure}

In Fig.~\ref{fig:parton_mom} (left) we present the percentage change for the $u, d$ and $s$ distributions as well as the gluon when QED kernels are included, against a default of pure QCD kernels at NNLO. The effect of QED evolution on the quarks, prior to refitting, is relatively modest, as expected due to the $\mathcal{O}(\alpha/\alpha_S)$ relevance of the QED splitting kernels in comparison to those of QCD. Although the change appears to grow at low $x$, this is in fact an artefact of the gluon PDF parameterisation, the expression for which is reproduced here for convenience: 
\begin{equation}\label{eq_gluon_2}
    xg(x,Q_0^2)= A_g(1-x)^{\eta_g}x^{\delta_g}\Bigg(\sum_{n=1}^{2}a_{g,i}T_i^{Ch}(y(x))\Bigg) + A_{g'}(1-x)^{\eta_{g'}}x^{\delta_{g'}}. 
\end{equation}
Here, the two competing contributions to this expression dominate the form of the input distribution (and therefore subsequent effects in the evolution of the sea) at low $x$. 
In particular, there is a strong correlation between the coefficients of the first and second terms, with the former term tending to increase the gluon at low $x$ during a fit, while the latter tends to decrease it.
A delicate balance and cancellation between these effects is seen to provide the best fit quality.
However, unlike the other parameters in this expression, $A_g$ is determined solely from the requirement that the momentum sum rule \eqref{eq_mom} be satisfied.
  If all other parameter values are taken from a fit using purely QCD kernels,
   the extra momentum provided by $x\gamma(x,Q_0^2)$ at input is compensated by a reduction of $A_g$, which diminishes the gluon contribution at low $x$. Such an effect disrupts the delicate cancellation between the terms described above.
This is seen to reduce the overall gluon momentum during the evolution, as well as that of the quark singlet distributions, as the latter at low $x$ are primarily driven by DGLAP emission from the gluon.
Therefore, a reduction in $g$ is expected and observed to have a knock-on effect in the same region, as shown in Fig. \ref{fig:parton_mom} (left).

In Fig.~\ref{fig:parton_mom}  (right) show the effect of on the quarks of refitting, described in more detail in Section~\ref{ssec_qed_fit}.
 We can see that the exaggerated effects of the evolution at low $x$ are compensated by the other parameters of the gluon, as discussed above. 
On the other hand, the behaviour of the partons at high $x$, which shows a small reduction in the singlet distributions are a genuine effect due to the inclusion of the QED contribution to $P_{qq}$. In particular, this reduction is primarily a natural consequence 
of the $q\rightarrow q+\gamma$ emission,
which at high $x$ has the effect of reducing the quark singlet momenta, with corresponding increases in $x\gamma(x,Q^2)$. 

We note that although the $s$ distribution experiences a larger magnitude of change due to QED than that of the other partons, this effect is a consequence of the $s+\bar{s}$ distribution being less well constrained by the data, and therefore more sensitive to the effects of refitting, rather than having an enhanced sensitivity to the effects of QED. 

\begin{figure}
\scalebox{.6}{\input{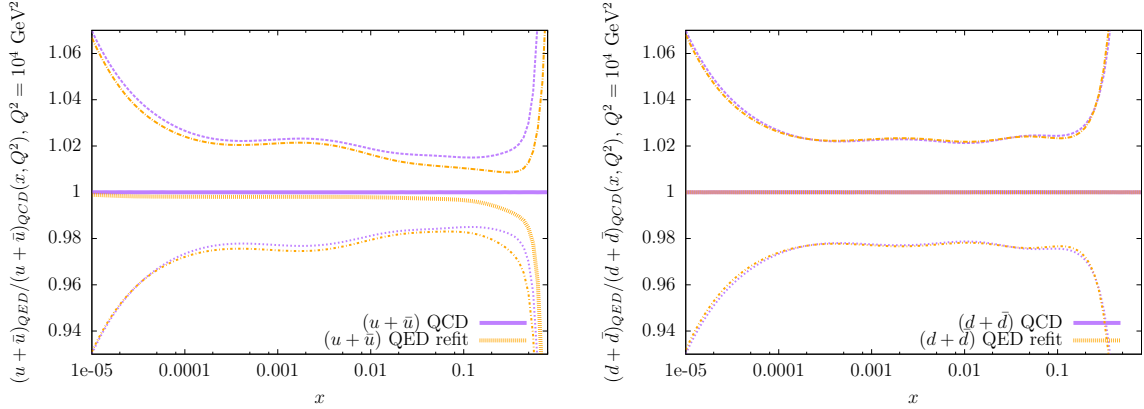}}
\begin{raggedright}
\scalebox{.6}{\input{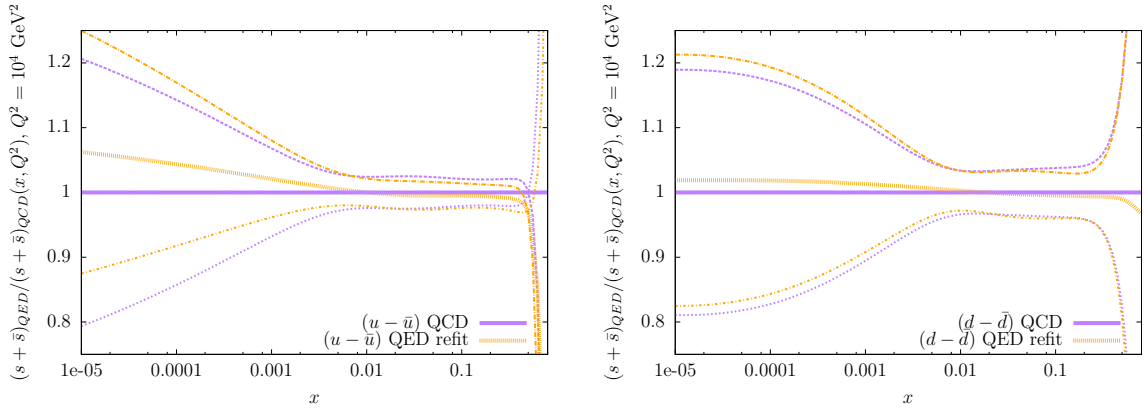}}
\end{raggedright}
\caption{The ratio of the $(u+\bar{u})$, $(d+\bar{d})$ distributions (with uncertainties) fit with and without the effects of QED in the evolution (both at NNLO in QCD) at $Q = 100$ GeV.}
\label{fig:uplus_dplus}
\end{figure}

In Figs. \ref{fig:uplus_dplus} - \ref{fig:uv_dv} we show the ratio of the PDFs with and without QED effects, including the corresponding PDF uncertainties. 
We can see that upon refitting the singlet ($q+\bar{q}$) and gluon PDFs all lie within the PDF uncertainties of the pure QCD fit, 
with the central values and uncertainties remaining only modestly affected,
with $\mathcal{O}(2\%)$ reduction for the $s+\bar{s}$ distribution, (with a slight increase in the reduction at high $x$, due to the effect of QED splittings mentioned above).
The up valence quark, $u_V$ and to a lesser extent the down valence quark $d_V$, are most sensitive to QED effects, with a $\mathcal{O}(2-5\%)$ change at low $x$ in their central values, though this is relatively marginal given the large uncertainties ($\sim 20\%$) in the valence quark PDFs in this region.

\begin{figure}
\scalebox{.6}{\input{SPLUS_REFIT.tex}}
\begin{raggedright}
\scalebox{.6}{\input{GLUON_REFIT.tex}}
\end{raggedright}
\caption{The ratio of the $(s+\bar{s})$, $g$ distributions (with uncertainties) fit with and without the effects of QED in the evolution (both at NNLO in QCD) at $Q = 100$ GeV.}
\label{fig:splus_g}
\end{figure} 

\begin{figure}
\scalebox{.6}{\input{UV_REFIT.tex}}
\begin{raggedright}
\scalebox{.6}{\input{DV_REFIT.tex}}
\end{raggedright}
\caption{The ratio of the $(u-\bar{u})$, $(d-\bar{d})$ distributions (with uncertainties) fit with and without the effects of QED in the evolution (both at NNLO in QCD) at $Q = 100$ GeV.}
\label{fig:uv_dv}
\end{figure} 

\begin{figure}[h]
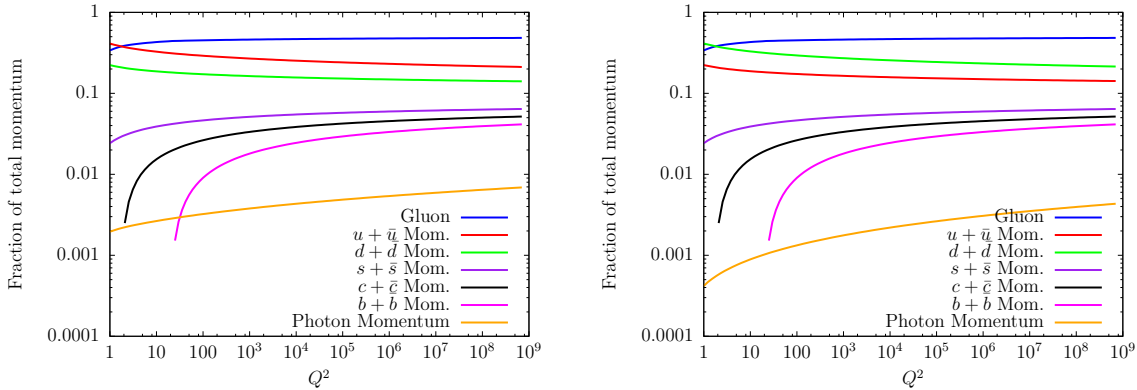

\scalebox{.6}{\input{momentum_Q2.tex}}
\begin{raggedright}
\scalebox{.6}{\input{momentum_Q2_neut.tex}}
\end{raggedright}
\caption{The percentage of the momentum carried by the  partons, including the photon PDF, in the proton (left) and neutron (right) as a function of $Q^2$ when QED evolution is included.}
\label{fig:photon_mom}
\end{figure}

In Fig.~\ref{fig:photon_mom} we see the details of the momentum carried by 
each of the partons as a function of $Q^2$ for both the proton and neutron. 
At input the fractional momentum carried by the photon in the proton is 
$0.00196$, and this increases to about $0.007$ at very high $Q^2$. In the 
neutron the input figure is much smaller, i.e. $0.0003$, but the rate 
of increase at higher $Q^2$ is comparable to the proton, though a little lower
due to the dominant radiation at high $x$ being from down quarks rather than up quarks. 

\begin{figure}
\centerline{%
\scalebox{.65}{\input{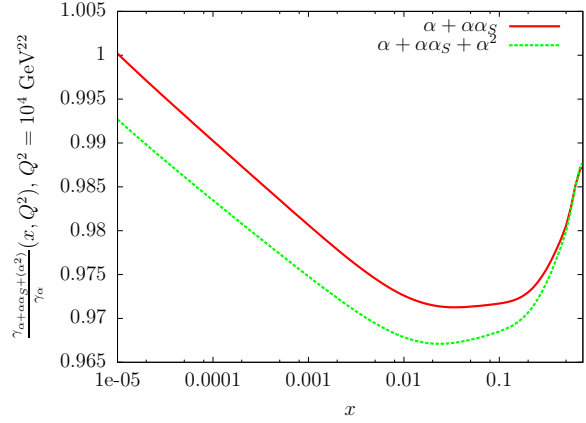}}} 
\caption{Ratio of the Photon PDF with and without $\mathcal{O}(\alpha\alpha_S)$,$\mathcal{O}(\alpha^2)$ corrections, at $Q^2 = 10^4$ GeV\textsuperscript{2}.}
\label{fig:alpha}
\end{figure}

\begin{figure}
\centerline{%
\scalebox{.65}{\input{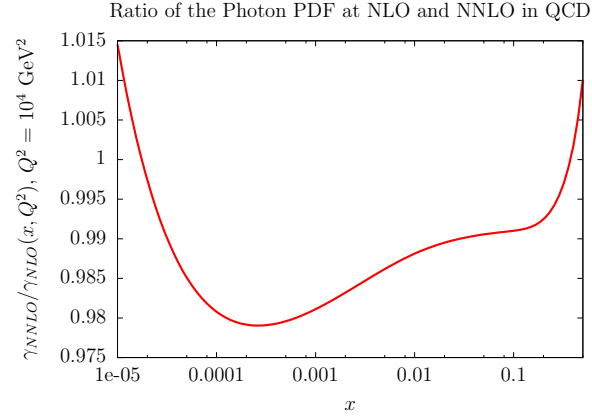}}}
\caption{Ratio of the Photon PDF at NLO and NNLO in QCD during DGLAP evolution, at $Q^2 = 10^4$ GeV\textsuperscript{2}.}
\label{fig:NLO_VS_NNLO}
\end{figure}

In Fig.~\ref{fig:alpha} we show the effect of the higher and mixed order corrections to the evolution on $x\gamma(x,Q^2)$.
We can see that the $\mathcal{O}(\alpha\alpha_S)$ and $\mathcal{O}(\alpha^2)$ kernels are seen to reduce the photon distribution by $\sim 1-3\%$, particularly at high $x$. The effect induced by the $\mathcal{O}(\alpha^2)$ kernels is of $\mathcal{O}(0.5-1\%)$,
 and further changes associated with the exclusion of yet higher orders in perturbation theory are expected to be even smaller. 
Since other sources of uncertainty, discussed in Section \ref{ssec_unc} are somewhat larger, it is not thought that such scale uncertainties will be significant for the photon at the level of accuracy being discussed in this paper.
Similarly, as shown Fig. \ref{fig:NLO_VS_NNLO}, the QCD order of the DGLAP evolution is found to have a modest effect on the resultant photon PDF produced. The photon experiences a slight reduction for intermediate values of $x$, $\mathcal{O}(1-2\%)$, with a slight increase at high and low $x$. This is largely due to differences in the underlying quark singlet, which as previously noted, have a strong role in influencing the form of $x\gamma(x,Q^2)$ at higher scales.

\subsection{Results of Global fits with QED corrections}\label{ssec_qeda_fit}

The fitting procedure is broadly similar to  that of MMHT14
, with the exception of changes to the structure function fits described below, and the inclusion of some new data as described in Section~\ref{ssec_qcd_basis}. In Section \ref{sec_pheno} we detail the results of an alternative fit that also includes high mass Drell-Yan data, which is seen to have some sensitivity to the inclusion of QED effects.

\subsubsection{The quality of the global fits}

\begin{table}[]
    \centering
\begin{tabular}{ |p{3cm}|p{3cm}||p{3cm}|p{3cm}|  }
 \hline
 \multicolumn{4}{|c|}{Change in $\chi^2$ due to QED evolution compared to MMHT14+HERA I+II} \\
 \hline
 NLO before fit&NLO after fit &NNLO before fit&NNLO after fit\\
  \hline	
 4180 (+41) & 4151(+12) & 3574 (+42)& 3539 (+7)\\\hline
\end{tabular}
    \caption{The total $\chi^2$ for partons with the effects of QED, both prior to and after refitting the parton parameters, at NLO and NNLO. Before the fit, the parameters derived from the QCD fits described in Section~\ref{ssec_qcd_basis} are used. The NLO fit contains 3609 data points, while the NNLO contains a total of 3276 (since the later omit some jet data). The numbers in brackets show the change in $\chi^2$ due to the inclusion of the QED corrections.}
    \label{tab:chisquareds}
\end{table}

 In Table \ref{tab:chisquareds}, we provide the change in the total $\chi^2$ in the fit to all data after the inclusion of all QED effects at NLO and NNLO, before and after refitting the partons. The full breakdown between individual datasets is given in Table~\ref{tab:chibreak}. We can see that the change in fit quality due to both the changes in the evolution and $\mathcal{O}(\alpha)$ corrections to the structure functions detailed below are modest. While the effects purely driven by the evolution naturally lead to an increase in $\chi^2$ (where the pure QCD fit parameters are used), this is somewhat reduced after refitting the partons to data with the full QED effects included. However, some increase with respect to the original QCD fit is still observed after refitting. From Table~\ref{tab:chibreak} we can see that this is primarily due to tension with the BCDMS $F_2$ and ZEUS CC data, with the former responsible for a $\sim+6$ increase in the total $\chi^2$ and the latter $\sim+2$. This is somewhat compensated for by a $\sim-2$ reduction from a slightly improved fit to $F_{2}$ and $F_{3}$ data from the NuTeV experiment, which see a mild improvement to the fit.

\begin{table}[]
{\scriptsize\begin{tabular}{ |p{8.5cm}||p{3.5cm}|p{3.5cm}|  }
 \hline
 \textbf{Data set}&\textbf{$\chi^2/N_{pts}$ NNLO before fit (QCD)}&\textbf{$\chi^2/N_{pts}$ NNLO after fit(QCD+QED)}\\
  \hline	
 BCDMS $\mu p$ $F_2$ \cite{BCDMS_1} &  178 / 163 & 182 / 163 (+4)\\
 BCDMS $\mu d$ $F_2$ \cite{BCDMS_2} &  142 / 151 & 144 / 151 (+2)\\
 NMC $\mu p$ $F_2$ \cite{NMC_1}&  124 / 123 & 125 / 123 \\
 NMC $\mu d$ $F_2$ \cite{NMC_1}&  108 / 123 & 108 / 123 \\
 NMC $\mu n/\mu p$ $F_2$ \cite{NMC_2}&  128 / 148 & 127 / 148 \\
 E665 $\mu p$ $F_2$ \cite{E665}&  65 / 53 & 65 / 53 \\
 E665 $\mu d$ $F_2$ \cite{E665}&  61 / 53 & 61 / 53 \\
 SLAC $e p$ $F_2$ \cite{SLAC_1}\cite{SLAC_2}&  31 / 37 & 31 / 37 \\
 SLAC $e d$ $F_2$ \cite{SLAC_1}\cite{SLAC_2}&  26 / 38 & 25 / 38 \\
 NMC/BCDMS/SLAC/HERA $F_L$ \cite{NMC_1,BCDMS_1,SLAC_2,HERA_FL_1,HERA_FL_2,HERA_FL_3}&  66 / 57 & 66 / 57 \\\hline
 E866/NuSea $pp$ DY \cite{E866_1}&  224 / 184 & 223 / 184 \\
 E866/NuSea $pd/pp$ DY \cite{E866_2} &  11 / 15 & 11 / 15 \\\hline
 NuTeV $\nu N$ $F_2$ \cite{nutev}&  37 / 53 & 36 / 53 (-1)\\
 CHORUS $\nu N$ $F_2$ \cite{chorus}&  29 / 42 & 29 / 42 \\
 NuTeV $\nu N$ $xF_3$ \cite{nutev}&  31 / 42 & 31 / 42 \\
 CHORUS $\nu N$ $xF_3$ \cite{chorus}&  19 / 28 & 19 / 28 \\
 CCFR $\nu N \rightarrow \mu\mu X$ \cite{CCFR}&  77 / 86 & 78 / 86 \\
 NuTeV $\nu N \rightarrow \mu\mu X$ \cite{CCFR}&  42 / 40 & 41 / 40 \\\hline
 HERA I+II CC $e^+ p$ \cite{HERA_I_II} & 52 / 39 & 52 / 39  \\
 HERA I+II CC $e^- p$ \cite{HERA_I_II} & 63 / 42 & 65 / 42 (+2) \\
 HERA I+II NC $e^+p$ 920 GeV \cite{HERA_I_II} & 510 / 402 & 510 / 402 \\
 HERA I+II NC $e^-p$ 920 GeV \cite{HERA_I_II} & 239 / 159  & 240 / 159 (+1) \\
 HERA I+II NC $e^+p$ 820 GeV \cite{HERA_I_II} & 88 / 75 & 88 / 75 \\
 HERA I+II NC $e^-p$ 575 GeV \cite{HERA_I_II} & 261 / 259 & 262 / 259 \\
 HERA I+II NC $e^-p$ 460 GeV \cite{HERA_I_II} & 246 / 209 & 246 / 209 \\
 HERA $ep$ $F_2^{charm}$ \cite{F2C}&  80 / 52 & 80 / 52 \\
 D{\O} II $p\bar{p}$ incl. jets \cite{D0_1}&  117 / 110 & 117 / 110 \\
 CDF II $p\bar{p}$ incl. jets \cite{CDF_1} &  60 / 76 & 60 / 76 \\
 CDF II $W$ asm. \cite{CDF_2} & 16 / 13 & 15 / 13 \\
 D{\O} II $W \rightarrow \nu e$ asym. \cite{D0_2} &  31 / 12 & 30 / 12\\
 D{\O} II $W \rightarrow \nu \mu$ asym. \cite{D0_3}&  16 / 10 & 16 / 10 \\
 D{\O} II $Z$ rap. \cite{D0_4}&  17 / 28 & 17 / 28 \\
 CDF $Z$ rap. \cite{CDF_3}&  40 / 28 & 40 / 28 \\\hline\hline
 ATLAS $W^+$, $W^-$, $Z$ \cite{ATLAS-WZ}&  41 / 30 & 41 / 30 \\
 CMS $W$ asymm $p_T > 35$ GeV \cite{CMS-W}&  7 / 11 & 7 / 11 \\
 CMS asymm $p_T >$ 25 GeV, 30 GeV \cite{CMS-asym} &  8 / 24 & 8 / 24 \\
 LHCb $Z\rightarrow e^+ e^-$ \cite{LHCb-1}&  22 / 9 & 22 / 9 \\
 LHCb $W$ asymm $p_T > 20$ GeV \cite{LHCb-2} &  14 / 10 & 13 / 10 \\
 CMS $Z\rightarrow e^+ e^-$ \cite{CMS-Z}&  23 / 35 & 22 / 35 \\
 ATLAS high-mass Drell-Yan \cite{DY-ATLAS}&  17 / 13 & 18 / 13 \\
 CMS double diff. Drell-Yan \cite{CMS-DY} &  152 / 132 & 152 / 132 \\
 Tevatron, ATLAS, CMS $\sigma_{t\bar{t}}$* \cite{tt-1,tt-2,tt-3,tt-4,tt-5,tt-6,tt-7}&  14 / 18 & 14 / 18 \\ \hline \hline
 \textbf{All data} &  \textbf{3532 / 3276} & \textbf{3539/3276} (+7)\\
 \hline
\end{tabular}
}
\caption{The $\chi^2$ breakdown showing $\chi^2/N_{pts}$ by data set for NNLO QCD and NNLO QCD + QED PDF fits. }\label{tab:chibreak}
\end{table}

\subsubsection{The photon PDF of the proton}

\begin{figure}
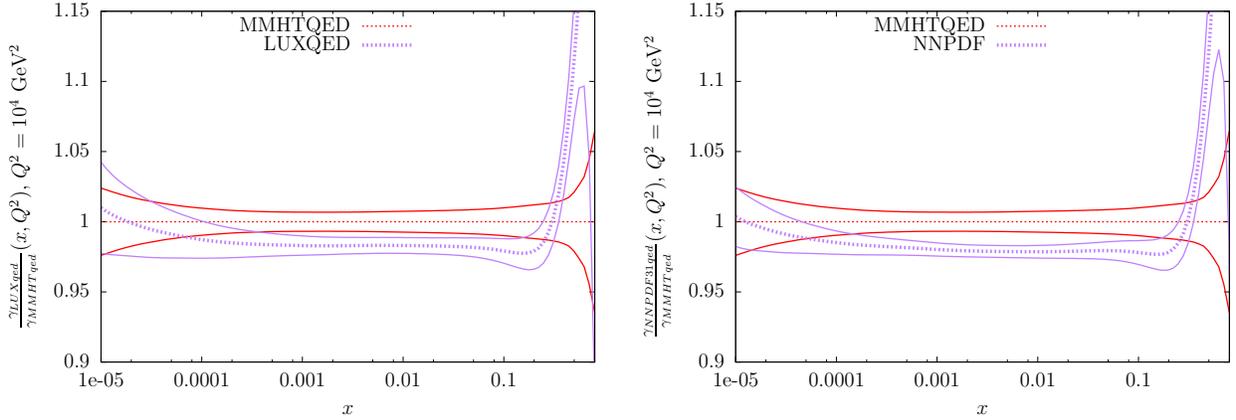

\scalebox{.65}{\input{PHOT_COMPARISON-LUX.tex}}
\begin{raggedright}
\scalebox{.65}{\input{PHOT_COMPARISON-NNPDF.tex}}
\end{raggedright}
\caption{The ratio of Photon PDFs between the LUXqed and NNPDF3.1luxQED sets with that of MMHT, at $Q^2 = 10^4$ GeV\textsuperscript{2}.}
\label{fig:comparison}
\end{figure}

\begin{figure}
\centerline{
\scalebox{0.7}{\input{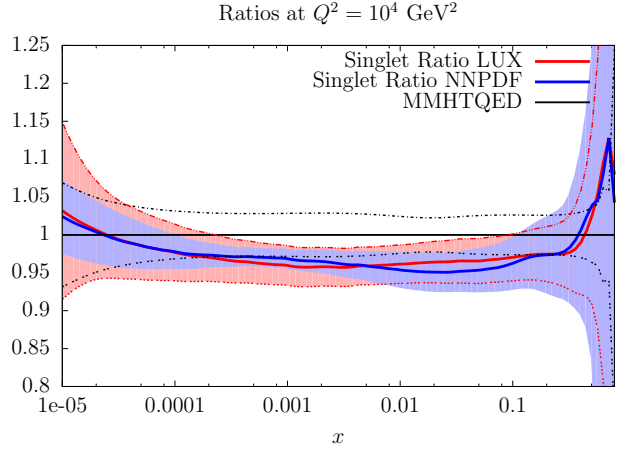}}}
\caption{The ratios of the charged singlet, $\sum_i e_{q_i}^2(q+\bar{q})$, between the LUXqed (which in turn adopts the quark and antiquark PDFs of PDF4LHC15) and NNPDF3.1LUXqed, against that of our set.}
\label{fig:luxsinglet}
\end{figure}

In Fig.~\ref{fig:comparison} we compare our photon PDF with those of LUXqed and NNPDF3.1luxQED, and the agreement is found to be quite good, i.e. they are within $\sim2\%$ over a broad range of $x$, diverging somewhat at high $x$ where uncertainties are seen to be large, and are close to the LUXqed photon. In particular the MMHT photon displays a very slight tendency to be somewhat larger in the intermediate range of $x$ and predicts a somewhat smaller photon at lowest $x$.
This is to some extent explained by the fact that the charge-weighted singlet ($\sum_{i}e_{q_i}^2(q+\bar{q})$) differs between MMHT2015qed and those of PDF4LHC15$\_$nnlo$\_$100 \cite{pdf4lhc} (which are the underlying partons used for LUXqed in the higher $Q^2$ representation of $F_{2,L}$) and NNPDF3.1 \cite{nnpdf2} as shown in Fig. \ref{fig:luxsinglet}. Note that the MMHT2015qed baseline PDFs are a percent or more larger than those of MMHT2014 at low $x$ and high-$Q^2$ as a consequence of fitting to the updated HERA data. This $\sim 2-4\%$ reduction of the charge weighted singlet between the sets as compared with ours in the range $10^{-4} < x < 10^{-1}$ then leads to a reduction in the relevant photon PDF ratios, as the evolution of $x\gamma(x,Q^2)$ is sensitive to this combination of partons. We also note that the largest discrepancy in the photon PDFs between MMHT2015qed and LUXqed at $x\sim 0.5-0.6$ is also a common feature of the charge-weighted quark difference in the relevant $x>0.5$ region.

\begin{figure}
\centerline{
\scalebox{0.7}{\input{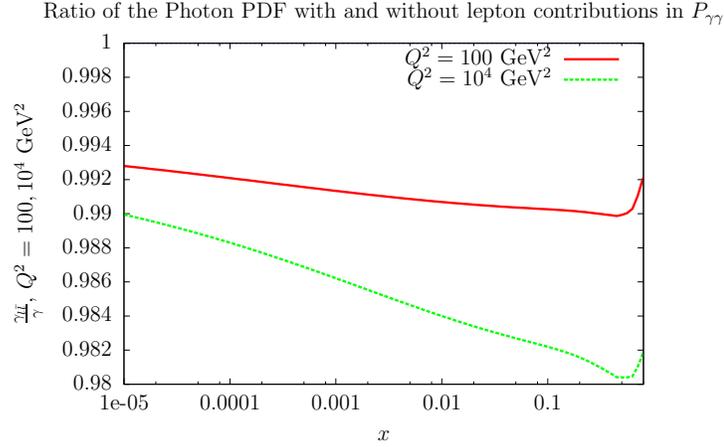}}}
\caption{The ratios of the photon with and without the sum over lepton
charge $\sum e_l^2$ contribution (the right hand side of eq. \ref{eq_leptsplit}) in the 
${\cal O}(\alpha)$ contribution to $P_{\gamma\gamma}$.}
\label{fig:lepton_rat}
\end{figure}

Another reason why we anticipate that the $x\gamma(x,Q^2)$ as outlined in this work may be somewhat greater in value, in an intermediate range in $x$, compared to that of LUXqed is due to the exclusion of lepton splitting contributions in our DGLAP evolution, which are included in the evolution used to develop the LUXqed set. In Section \ref{ssec_DGLAP} we explicitly neglected the sum over lepton charges in Eq. \ref{eq_leptsplit}. In general, since $\gamma\rightarrow l\bar{l}$ splittings should reduce the photon distribution (nearly uniformly since it occurs as a coefficient to $\delta(1-x)$ in $P_{\gamma\gamma}$), one expects that excluding this term should lead to a somewhat increased photon. To estimate the effect of including this term, in Fig. \ref{fig:lepton_rat} we draw a comparison to $x\gamma(x,Q^2)$ evolved with $\mathcal{O}(\alpha)$ lepton splittings included in evolution and as anticipated find that this term does lead to a $\mathcal{O}(1-2\%)$ reduction, which becomes more pronounced at higher $Q^2$. Along with the ratio of the charged singlet used in the evolution, neglecting lepton splittings\footnote{Note that excluding this term is still a reasonable approximation given that a fully consistent treatment with a coupled DGLAP evolution would require the development of lepton PDF distributions which as discussed in \cite{lepton_pdfs} are found to have a negligible impact on the evolution of the PDFs on the whole.} leads to an independent source of enhancement for our $x\gamma(x,Q^2)$, further accounting for the difference seen in Fig. \ref{fig:comparison}.

Common to all the sets are errors of $\mathcal{O}(1\%)$, displaying the remarkable improvements in accuracy seen in photon PDFs developed on the strategy outlined in this paper and that of \cite{lucian} and \cite{a}, in comparison to that of older sets. A full breakdown of the contributing sources of error are explored in Section~\ref{ssec_unc}.

\subsubsection{QED Corrected Structure Functions}

The PDFs are related to the measured structure functions by the standard formulae
 \begin{equation}\begin{split}
   F_i(x,Q^2) = x \sum_{q,\bar{q}} e_q^2 \int_x^1 \frac{dz}{z}q(z,Q^2)\Big\{\delta\left(1-\frac{x}{z}\right)+\frac{\alpha_S}{2\pi}C_{i,q}\left(\frac{x}{z}\right)+...\Big\} + \\
    x \sum_{q,\bar{q}} e_q^2 \int_x^1 \frac{dz}{z}g(z,Q^2)\Big\{\frac{\alpha_S}{2\pi}C_{i,g}\left(\frac{x}{z}\right)+...\Big\}.
\end{split}
\end{equation} 
where $i$ ($=2,3,L...$) labels the structure function and $C_{q,g}$ are the corresponding coefficient functions. The introduction of a photon PDF and of QED corrections to the DGLAP splitting kernels requires that we also modify the expression for these to include $\mathcal{O}(\alpha)$ corrections, in particular introducing terms of the form $C_{\gamma}^{(\alpha)}\otimes\gamma(z,Q^2)$, for both Neutral Current (NC) and Charged Current (CC) processes. 

\begin{figure}
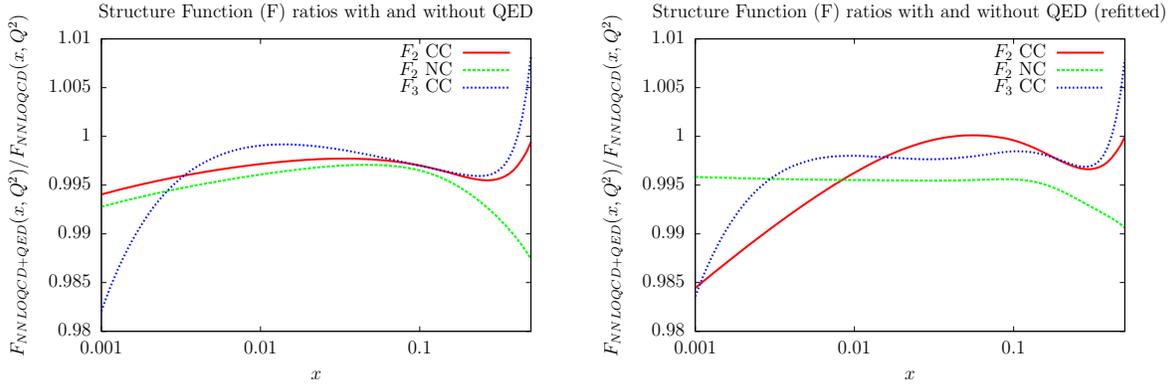

\scalebox{.6}{\input{SF_Ratio_unfit.tex}}
\begin{raggedright}
\scalebox{.6}{\input{SF_Ratio_fit.tex}}
\end{raggedright}
\caption{The ratio of the Charged and Neutral Current $F_2$ and Charged Current $xF3$ for the proton, with and without the effects of QED, both at $Q^2 = 10^4$ GeV\textsuperscript{2}. (Left) the effects of na\"{i}ve inclusion of QED splitting kernels without the refitting of the partons (in which the artificial reduction in the low $x$ gluon and hence the sea quarks has an enhanced effect, as discussed in the text). (Right) The ratio of Structure Functions after refitting the partons, with modest effects observed in $F_2$ CC and NC.}
\label{fig:structure}
\end{figure} 

In Fig.~\ref{fig:structure} we show the effect of these changes with and without refitting. Again, the sensitivity introduced by the gluon parameterisation is seen to have an effect at low $x$, reducing $F_{2,3}$ somewhat, while after fitting, the CC structure functions $F_{2,3}$ are moderately decreased at low $x$. In the NC case however, $F_2$ is generally reduced by $\mathcal{O}(0.5\%)$, as anticipated by the fact that the introduction of QED in the evolution is seen in general to diminish the quark singlet content, see Fig.~\ref{fig:parton_mom}.

\subsubsection{Effects of QED on $\alpha_S$ determination in the global PDF fit}

In addition to the fit described above, we have also performed 
a simultaneous fit to the strong coupling 
, $\alpha_S(M_Z)$.
The value typically used during the evolution and the comparison to data is taken as a fixed value $\alpha_S(M_Z) = 0.118$, which reflects a combination of both the best fit value exclusively from our fit to data, 
 and the independent inclusion of the world average of $\alpha_S(M_Z) = 0.1181 \pm 0.0011$ \cite{PDG}, as discussed in Section 5.1 of \cite{MMHT}.

In principle, one might expect that the value of $\alpha_S(M_Z)$ found after refitting with the effects of QED included will be somewhat less than that in a pure QCD fit. This is because at leading order, the effect on the $q+\bar{q}$ distributions during the evolution, particularly at high $x$, is due to gluon emission, $q\rightarrow q g$, which leads to a slight reduction of the singlet. In a pure QCD fit, the parameters that provide the best fit are a combination of both the input distribution and a value of $\alpha_S(M_Z)$ which drives gluon emission at a rate (determined by $P_{qq}^{(QCD)}$) in the evolution such that the PDFs at higher scales are best fit to the data.

At LO in QED however, the electromagnetic coupling $\alpha$ plays virtually the same role in the evolution of the singlet distributions, diminishing the high $x$ content due to photon emissions $q\rightarrow q \gamma$. Therefore at LO, one can consider the inclusion of QED as an enhancement to $P_{qq}$ with an increased effecting coupling:\begin{equation}\label{eq:modified_alpha}
    \alpha_S \rightarrow \alpha' = \Big(\alpha_S + \frac{e_q^2\alpha}{C_F}\Big).
\end{equation}
 In a fit that includes the coupling constants as free parameters, one expects that $\alpha'$, rather than $\alpha_S$ would tend towards a value that best models the loss of the singlet during evolution to emission (whether to a photon or gluon). Since $\alpha_S$ is the only free parameter in the fit (where we adopt the world best measurement value for $\alpha$ \cite{PDG}), one naturally expects the best fit value for $\alpha_S$ to be reduced to accommodate the modification in eq. \eqref{eq:modified_alpha}. Na\"{i}vely, one may expect the magnitude of this reduction to compensate for the magnitude of the modification term $e_q^2 \alpha/ C_F \sim 10^{-3}$. Though small, this is similar to the global fit uncertainty on $\alpha_S$, and the effects of QED may therefore be significant in its determination.

This was also investigated in the development of the original MRST QED set~\cite{mrstqed}, where it was found that despite the above considerations, between the pure QCD and QCD+QED fit, $\alpha_S(M_Z)$ remained essentially unchanged. The reason found for this 
was that 
the fit (especially the NMC and HERA data) preferred a larger value for the gluon at low $x$, which is sensitive to $\alpha_S(M_Z)$ since $d F_2/d \ln Q^2 \propto \alpha_S P_{qg}\otimes g(Q^2)$.
However, the momentum carried by the photon detracts from that carried by the small-$x$ gluon and  
as a result, the change to the gluon at small $x$ has a tendency to require a larger value of $\alpha_S(M_Z)$ than would otherwise be obtained. This pulls in a direction opposite to the 
reduction of $\alpha_S(M_Z)$ as described above, and reduces the magnitude by which one might anticipate a change after refitting with the effects of QED.

With the updated QED parton framework, we find that $\alpha_S(M_Z)$ experiences a reduction from 0.1181 in the pure NNLO QCD case to 0.1180 in the fit with QED, while at NLO the result is unchanged within the numerical precision of the fit. Although at NNLO this does represent a small reduction, in neither case is allowing $\alpha_S$ to be free seen to improve the total fits by any significant degree, with $\Delta\chi^2 < 1$. However, in future global fits, the inclusion of QED effects in the partons may come to be significant as the accuracy of such measurements are improved.

\subsubsection{Photon-photon luminosity}

\begin{figure}
\centerline{
\scalebox{0.7}{\input{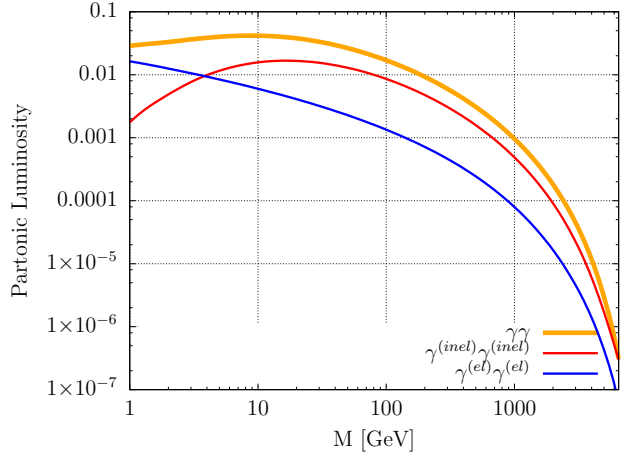}}}
\caption{$\gamma\gamma$ partonic luminosities as a function of invariant mass at Centre-of-Mass energies of 14 TeV. Note that for the elastic $\gamma^{(el)}\gamma^{(el)}$, the multi-particle interaction (MPI) effects are not included, and their inclusion would reduce (lower) the blue curve to some degree (discussed in text).}
\label{fig:luminosity}
\end{figure}

\begin{figure}
\centerline{
\scalebox{0.7}{\input{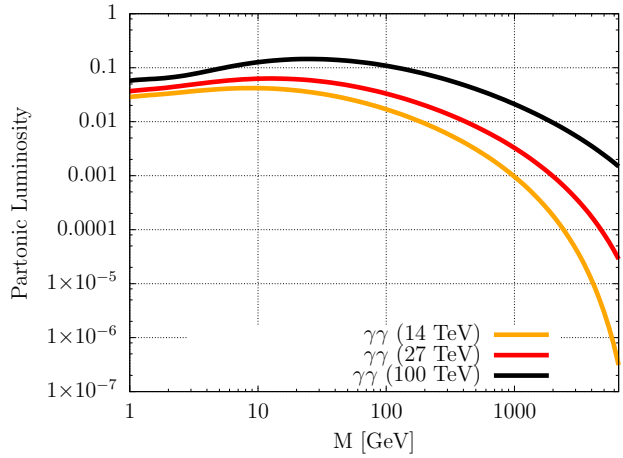}}}
\caption{$\gamma\gamma$ partonic luminosities as a function of invariant mass at Centre-of-Mass energies of 14, 27 and 100 TeV.}
\label{fig:luminosity_Q}
\end{figure}

A sense of the relevance of the photon PDF to particle production at colliders such as the LHC may be determined from an inspection of the $\gamma\gamma$ luminosity expected at these energies (14 TeV), shown in Fig.~\ref{fig:luminosity}. As seen in Fig.~\ref{fig:comparison}, our photon and that of other sets based on the LUXqed formulation show good agreement, and therefore our predicted $\gamma\gamma$ luminosity, ${\rm d} L_{\gamma\gamma}/{\rm d}\ln M^2$, bears a strong resemblance to others in the literature (see e.g. Fig. 19 in \cite{b}). Also shown in Fig. \ref{fig:luminosity_Q} is the expected luminosity for a High-Energy LHC proposal with (CoM) energy $\sqrt{s} = $27 TeV, and a Future Circular Collider with $\sqrt{s} = $100 TeV, where the total $\gamma\gamma$ luminosity is comparable to that of $\Sigma_i (q_i\bar{q}_i+\bar{q}_iq_i)$ at present LHC CoM energies (14 TeV).

Furthermore, as our photon PDF is separable by its elastic and inelastic components, we are able to distinguish between $\gamma^{(inel)}\gamma^{(inel)}$ and $\gamma^{(el)}\gamma^{(el)}$ contributions to the overall luminosity. The latter is of particular interest in the context of photon-initiated central exclusive production  (CEP -- see e.g.~\cite{N.Cartiglia:2015gve,Harland-Lang:2018iur}). In this process the protons collide peripherally, exchanging only photons while remaining in tact, such that they can be detected and their kinematic properties reconstructed in dedicated proton tagging detectors installed in association with ATLAS~\cite{exclusive_2} (AFP) and CMS~\cite{exclusive_3} (CT--PPS).

The cross section for this CEP process can be calculated within the so--called Equivalent Photon Approximation~\cite{EPA}, in which the photon flux associated with the colliding beam of charged particles may be expressed in terms of the elastic structure functions $F_{2,L}^{(el)}$, in a manner similar to that considered in this paper. The $\gamma^{(el)}\gamma^{(el)}$ luminosity, represented in Fig. \ref{fig:luminosity}, corresponds to precisely the luminosity that could be delivered in this approach.

However, this interpretation must be qualified with an important caveat, which is that for an exclusive production process, where both protons remain intact after scattering, one needs to multiply the final result obtained from the na\"{i}ve use of $\gamma^{(el)}$ as an incoming parton by a `soft survival' factor, corresponding to the probability of no additional particle production due to multi-particle interactions (MPI) \cite{survival}. 
Furthermore, the luminosities shown in Fig. \ref{fig:luminosity} can not directly be applied to the calculation of cross sections for more exclusive final states, such as when explicit cuts are placed on the presence of additional tracks within the central portion of the detector, but require suitable modification as in~\cite{Harland-Lang:2016apc}.

\begin{figure}
\centerline{
\scalebox{0.7}{\input{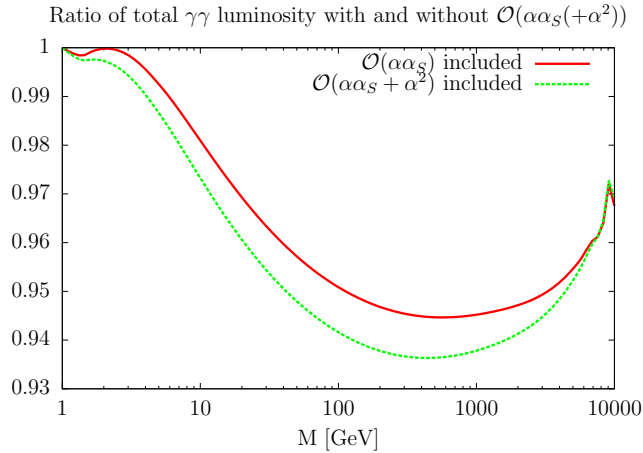}}}
\caption{Ratio of $\gamma\gamma$ partonic luminosity as calculated from $x\gamma(x,Q^2)$ with and without $\mathcal{O}(\alpha\alpha_S)$ and $\mathcal{O}(\alpha^2)$ DGLAP splitting kernels during evolution as a function of invariant mass, at a proton-proton Centre-of-Mass energy of 13 TeV.}
\label{fig:luminosity_alpha}
\end{figure}

We conclude this section with a discussion of the effect of higher and mixed orders of QED, $\mathcal{O}(\alpha\alpha_S)$ and $\mathcal{O}(\alpha^2)$ during the evolution and the significance of their impact on the total luminosity at present CoM energies at the LHC. As previously observed in Fig.~\ref{fig:alpha}, the inclusion of these higher order splitting functions in the evolution of $x\gamma(x,Q^2)$ have a tendency to reduce its magnitude, particularly at the higher range in $x$.
In Fig. \ref{fig:luminosity_alpha}, the proportional effects of such changes in $dL_{\gamma\gamma}/d\ln M^2$ are shown. We see that, above the electroweak and near TeV scales, the importance of these higher orders become significant, inducing a $\mathcal{O}(5\%)$ reduction in the total $\gamma\gamma$ luminosity.

\subsection{Uncertainties on the photon PDF}\label{ssec_unc}

Our  treatment of the contributions to the photon PDF uncertainty are in some cases identical to LUXqed.
However, as discussed in Section~\ref{ssec_DGLAP}, due to the lower starting scale adopted in our evolution procedure, we also include higher twist corrections in the form of a renormalon model, for which the undetermined coefficient $A'_2$ in eq. \eqref{eq_renormalon} is fit to the data, introducing an independent source of uncertainty. 

\begin{figure}
\centerline{
\scalebox{0.7}{\input{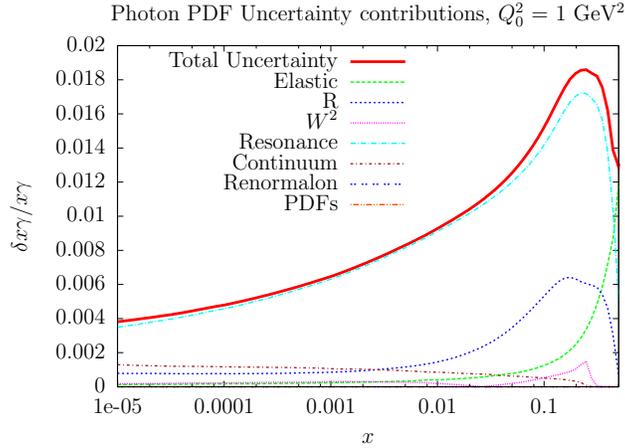}}}
\caption{Photon PDF Uncertainty contributions (added in quadrature to give the total uncertainty), $Q_0^2 = 1$ GeV\textsuperscript{2}. Note that the upper $x$ range has been restricted in this plot due to the effect of the kinematic cut given in \eqref{eq_xcut}}
\label{fig:uncertainties-1}
\end{figure}

For completeness, a full description of the uncertainty contributions is given below. 
The  size of the different sources of uncertainty as a function of $x$ and for different scales $Q^2$ is shown in Figs.~\ref{fig:uncertainties-1}-\ref{fig:uncertainties-100}.

\begin{figure}
\centerline{
\scalebox{0.7}{\input{ALL-UNCERTAINTIES-sqrt10.tex}}}
\caption{Photon PDF Uncertainty contributions (added in quadrature to give the total uncertainty), $Q^2 = 10$ GeV\textsuperscript{2}.}
\label{fig:uncertainties-10}
\end{figure}

\begin{figure}
\centerline{
\scalebox{0.7}{\input{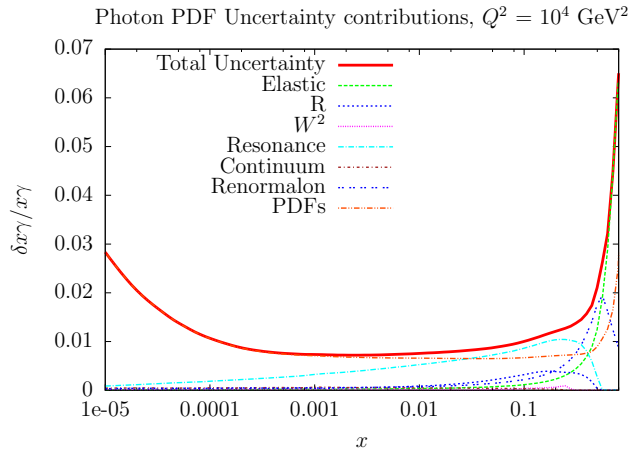}}}
\caption{Photon PDF Uncertainty contributions (added in quadrature to give the total uncertainty), $Q^2 = 10^4$ GeV\textsuperscript{2}.}
\label{fig:uncertainties-100}
\end{figure}

\begin{itemize}

    \item \textbf{Elastic:} The uncertainty contributions from the A1 fit for $F_2^{(el)}$ are twofold. In particular, the fits provided by A1 are given in the unpolarized and polarised forms, where the latter accounts for potential two photon exchange (TPE) processes between the lepton probe and the proton in DIS experiments. Following the approach of LUXqed, we use the latter for our estimate precisely because it provides constraints on TPE. As well as the intrinsic uncertainty provided by the A1 collaboration for this fit $\delta (F_2^{(el)})_a$, similarly to LUX, we adopt the symmetrised difference between the polarised and unpolarized fit as an independent source of error, $\delta (F_2^{(el)})_b$. The total uncertainty on $F_2^{(el)}$ is then simply given by the sum of these two contributions in quadrature.
    
    \item \textbf{R:} The contributions from $F_{L}$ are modelled in precisely the same manner as that of LUXqed, using the parameterisation of the form:\begin{equation}
        F_L(x,Q^2) = F_2(x,Q^2)\Big(1+\frac{4m_p^2x^2}{Q^2}\Big)\frac{R_{L/T}(x,Q^2)}{1+R_{L/T}(x,Q^2)}, 
    \end{equation} where $R_{L/T} = \sigma_L(x,Q^2)/\sigma_T(x,Q^2)$ represents the ratio between the absorption cross sections for longitudinal and transversely polarised photons. Our expression for this ratio is provided by the LUXqed group, who, following the procedure used by the HERMES collaboration \cite{e}, in turn adapt the expression from the $R_{1998}$ fit\cite{R} provided by the E143 Collaboration for use in low $Q^2$ regions and assign it a conservative $\pm 50\%$ uncertainty, which we also adopt.
    
    \item\textbf{W\textsuperscript{2}:} 
    As discussed in Section~\ref{ssec_input},
    two distinct fits for $F_2^{(inel)}$ are used above (HERMES \cite{e}) and below (CLAS\cite{d} and Cristy-Bosted\cite{f}) a threshold of $W_{cut}^2 = 3.5$ GeV\textsuperscript{2}. Since $W_{cut}^2$ is defined somewhat arbitrarily and theoretically induces some small amount of discontinuity in the contributions to $\gamma^{(inel)}$, we treat the cut value as an independent source of uncertainty, varying it the region $3 < W_{cut}^2 < 4$ GeV$^2$. Even with this relatively conservative approach, the uncertainty on $W_{cut}^2$ is seen to be vastly dominated by other sources.
    
    \item\textbf{Resonance:} The uncertainty of $F_2^{(inel)}$ in the resonance region is taken as the symmetrised difference between the CLAS fit, which is used as the standard for our input, and that of the Cristy-Bosted, similar to the procedure used by LUXqed.
    
    \item\textbf{Continuum:} The uncertainty of $F_2^{(inel)}$ in the continuum region is adapted directly from the uncertainty bands of the GDP-11 fit provided by the HERMES collaboration. This is a different type of uncertainty estimate 
from this source as that adopted by LUXqed, who vary the scale at which 
$F_2$ goes from being described by the GDP-11 fit to calculated in terms of 
the PDFs. However, each estimation of uncertainty is very small.

    \item\textbf{Renormalon:} For the fitting and uncertainty of the coefficient $A'_2$ in eq. \eqref{eq_renormalon}, we implemented the original renormalon model of \cite{i} into the calculation of the structure functions themselves, $F_{2,3}$, as used in the fit . $A'_2$ was then varied to induce a $\Delta \chi^2 = \pm 10$ change in the overall fit quality of the partons (as seen in Fig. \ref{fig_A_2} in Section~\ref{ssec_DGLAP}), creating a generous uncertainty band of $-0.4 < A'_2 < -0.2$, with a best fit value of -0.3. We note that our global fit to the data favours a renormalon contribution $\sim50\%$ greater than the value used in the original model by Dasgupta and Webber \cite{i}. At high $x$, this is seen to be a comparable source of uncertainty with that of $\delta(F_2^{(el)})$. Unlike all other terms discussed so far, the uncertainty in $A'_2$ enters during the evolution, rather than at input. 
    
    \item\textbf{PDFs:} Above the input scale $Q_0^2 = 1$ GeV\textsuperscript{2} the $\gamma^{(inel)}$ contributions are modelled solely from the splittings of other partons during the DGLAP evolution. Hence, the intrinsic uncertainty on the other PDFs propagate into the form of the photon PDF as it evolves. This reflects the standard 50 eigenvector uncertainties associated with the fit of the free parameters in the MMHT parameterisation (see eqs. \eqref{eq_parameter} and \eqref{eq_gluon}), which generate the uncertainty bands for all flavours of parton ($q, \bar{q}, g$), naturally generating uncertainties in the photon during splittings of the form $q\rightarrow q\gamma$ and $g\rightarrow q\bar{q}\gamma$. At low $x$, as is the case of LUXqed, this  dominates as the primary source of uncertainty.

\end{itemize}

Since our $\gamma^{(inel)}$ is evolved from a common starting scale, 
 we are alleviated of the consideration of matching scales between the photon and other partons (though this is seen to be negligible even when necessary, as shown for (M) in Fig.~15 of \cite{b}). Furthermore, in comparison to that of LUXqed, our set neglects certain contributions to the photon uncertainty. In particular, rather than the Twist-4 uncertainties considered by LUXqed for $F_L$ (which an inspection of (T) in Fig.~15 of \cite{b} reveals to be overwhelmingly dominated by other sources), our treatment of the Higher-Twist (HT) corrections to the structure function in the form of the renormalon lead to a more significant uncertainty at high $x$, consistent with our choice of a lower starting scale for the evolution.

Indeed, since our starting scale is at $Q_0^2 = 1$ GeV, as shown in Fig.~\ref{fig:uncertainties-1}, the uncertainties at input have a markedly different form to the kind that arises during the evolution. Naturally, effects that pertain to the evolution, (the PDF eigenvector uncertainties and the renormalon) are absent at this scale, and the dominating effects are seen to be the uncertainty on the resonance contribution to $F_2^{(inel)}$, the uncertainties on the Sachs form factors provided by the GD-11 fit ($\delta F_2^{(el)})$ and the uncertainty on $R_{L/T}$. As the evolution occurs however, the PDFs overwhelmingly dominate as the source of uncertainty at low $x$, and in conjunction with the uncertainty on the renormalon parameter $A'_2$, become significant contributions along with those of the Sachs form factors at higher $x$. 

It is noted that we do not account for the uncertainty that arises from the Higher Order (HO) terms missing from the QCD components of the evolution, as estimated in LUXqed. Although we have given an indication of the magnitude of the change in order from QCD (from NLO to NNLO) in the evolution in Fig. \ref{fig:NLO_VS_NNLO} of the previous section (which broadly corresponds to the (HO) band in fig. 15 of \cite{b}), we do not treat this difference as an independent source of uncertainty, since PDFs have typically been provided at both NLO and NNLO in QCD, each with independently derived uncertainty bands. 
Despite not being included as a default, recent work \cite{HO} has begun to explore the possibility of incorporating such uncertainties into the PDF fitting framework of MMHT in a standard manner.

Overall, we note the similarity between the form of our uncertainty with others, being less than 2\% for $10^{-5} < x < 0.5$, demonstrating a drastic improvement with early photon PDF sets such as MRST2004QED \cite{mrstqed} and NNPDF2.3 \cite{nnpdf1}.

We provide the photon PDF along with the quark, antiquark and gluon PDFs in 
grids which also contain all information about the uncertainties. PDF sets are 
typically provided as grids in the LHAPDF6 format, with each grid 
representing either the central value of the PDFs, or the PDFs at a 
given $\pm$ eigenvector direction in the independent parameter space PDFs. 
As noted above, as well as the uncertainties that are routinely given in such 
sets associated with the non-photon PDF parameters, the set that is produced 
as a result of the work described here now contains uncertainties 
associated with the photon parameters at input and the $A'_2$ parameter for 
the renormalon in the evolution. The grids will be discussed in more detail 
in the Appendix.

\section{High Mass Drell-Yan} \label{sec_pheno}
\subsection{QED and Photon PDF sensitivity in High Mass Drell-Yan}\label{ssec_HMDY}

\begin{figure}[h]
\scalebox{.6}{\includegraphics{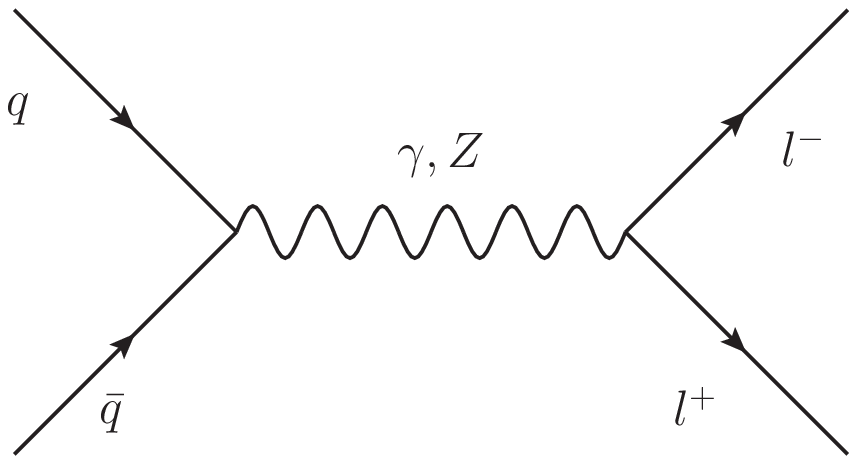}}
\scalebox{.6}{\includegraphics{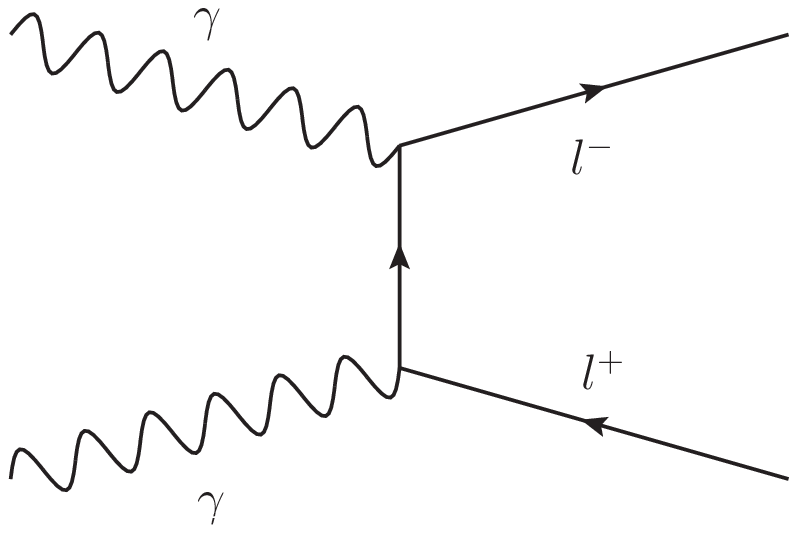}}
\scalebox{.6}{\includegraphics{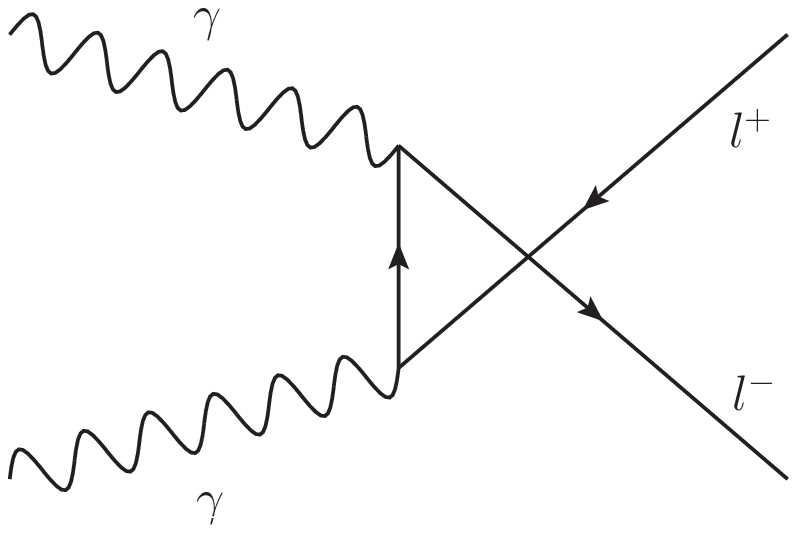}}
\caption{Leading order Drell-Yan production (left), with diagrams (centre, right) indicating $\mathcal{O}(\alpha)$ photon-initiated (PI) contributions to the total cross section.}
\label{fig:DY_processes}
\end{figure}

In order to explore the phenomenological implications of our photon PDF set, we calculate the effects on the double differential cross section for lepton pair (Drell-Yan) production at the LHC. This process is of particular interest, since the effects of QED, especially in the partons, is expected to be of non-negligible significance, particularly due the inclusion of $x\gamma(x,Q^2)$ as a contribution to the cross section.
Below, we will consider the impact of both including QED effects in the evolution of the PDFs as well as the addition of photon-initiated (PI) contributions, as shown in Fig.~\ref{fig:DY_processes}, where the photon PDF enters as a direct input for the colliding partons.

\subsection{Comparison with ATLAS Drell-Yan data}

In order to gauge the magnitude (and phenomenological significance) of these effects we compare to data provided by the ATLAS collaboration \cite{DY-ATLAS-2} for high mass (116 GeV $< m_{ll} <$ 1500 GeV) Drell-Yan lepton pair production. The focus on production at high mass is chosen in order to reduce the effects of the $Z$ production peak, $Q \sim M_Z = 91$ GeV, since the relative contribution of the PI processes are greater in the regions dominated by the $\gamma$ channel. Therefore, the effects of PI contributions are anticipated to be more readily observable at low, $m_{ll} \ll M_Z^2$, or high, $m_{ll} \gg M_Z^2$, lepton pair invariant masses. 

\begin{figure}
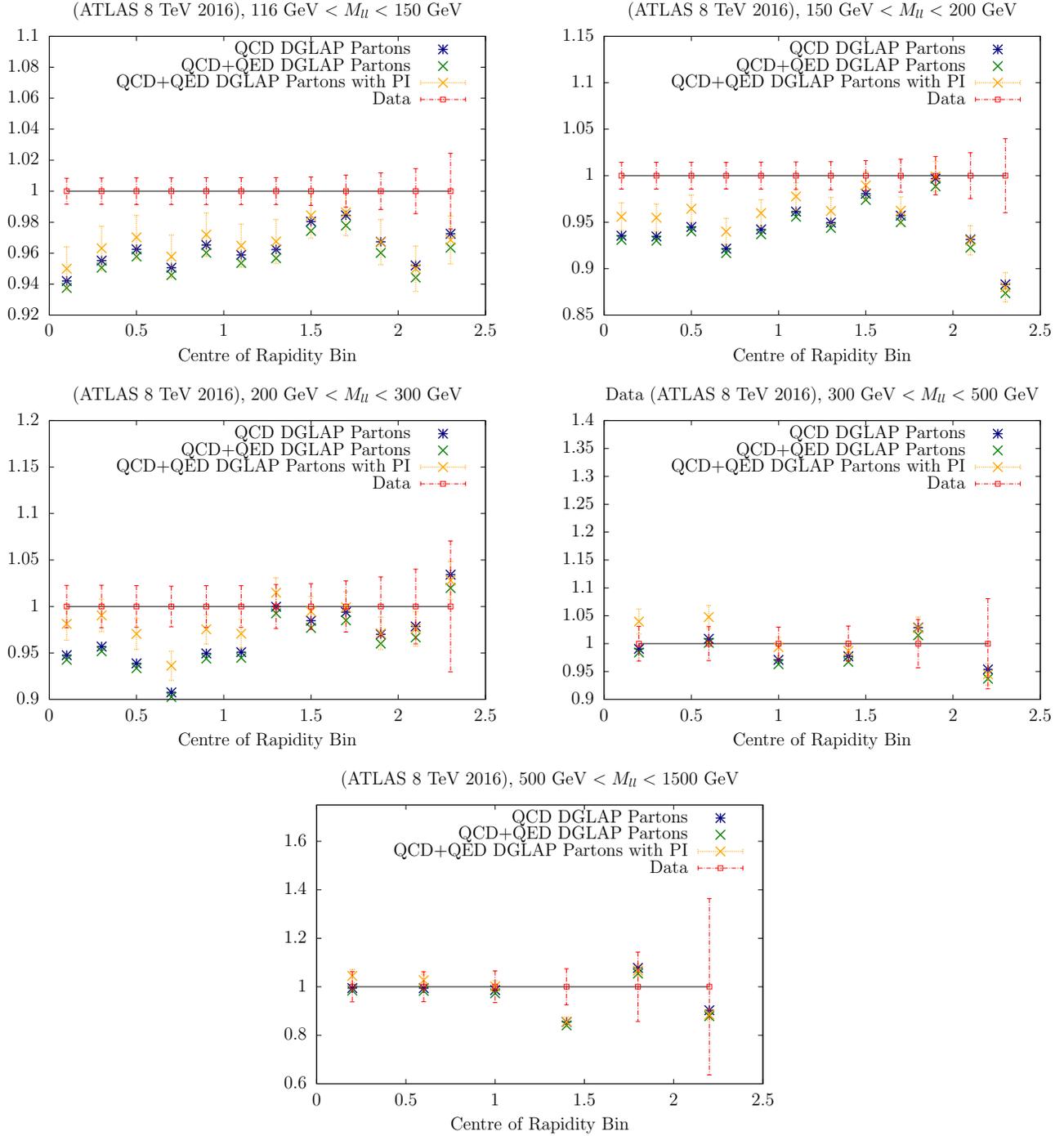

\centerline{%
\scalebox{0.7}{\input{theory_0_data_rat.tex}}
\scalebox{0.7}{\input{theory_1_data_rat.tex}}}
\centerline{\scalebox{0.7}{\input{theory_2_data_rat.tex}}
\scalebox{0.7}{\input{theory_3_data_rat.tex}}}
\centerline{\scalebox{0.7}{\input{theory_4_data_rat.tex}}}
\caption{The theory prediction/Data Ratio for ATLAS 8~TeV Drell-Yan data 
as a function of rapidity in different mass bins. Shown are the predictions 
using QCD only, QED included in PDF evolution but photon-initiated processes
not included, and full QCD plus QED including photon-initiated processes.}
\label{fig:DY_0}
\end{figure}

ATLAS provides double differential cross section measurements in 5 bins of lepton pair invariant mass, $m_{ll}$ and 12 or 6 pseudo-rapidity $\eta$ bins, depending on the mass region. Fig.~\ref{fig:DY_0} shows 
as a comparison for a range of cases:
 (a) a standard QCD fit partons at NNLO as outlined in Section~\ref{ssec_qcd_basis}, (b) with QED modified partons to provide cross section calculations at NNLO in QCD and (c) with QED modified partons and additional contributions to the cross section from $\mathcal{O}(\alpha)$ photon initiated processes as shown in Fig.~\ref{fig:DY_processes}. 

To calculate cross sections, we use grids provided by the xFitter collaboration \cite{xfitter}, at NLO in QCD (generated with MadGraph5$\_$aMC@NLO \cite{madgraph}, aMCfast \cite{amcfast} and FEWZ \cite{fewz}), and including PI processes at LO in QED. NNLO QCD corrections are included via $K$-factors.
Such grids were developed and used in \cite{xfitter} with the aim of determining $x\gamma(x,Q^2)$ from the same ATLAS data. These are then interfaced with a modified version of APPLgrid that we have adapted to include $\gamma\gamma$ processes for the final calculation.

In the following analysis it is emphasised that the contributions of PI processes implemented in the comparison to data will be most sensitive to $x\gamma^{(inel)}(x,Q^2)$, due to the prevalence of this contribution in comparison to $x\gamma^{(el)}(x,Q^2)$ at higher scales (as was seen in the lower part of Fig.~\ref{fig:elastic_rat} in Section~\ref{ssec_split}).

First, it is observed that the addition of QED in the process of DGLAP leads to a tendency to decrease the dominantly $q\bar{q}$ contribution to the cross section, increasingly so at higher rapidity. This is expected, as from Fig. \ref{fig:parton_mom} one observes that the quarks experience a reduction at high $x$ of $\sim1\%$ due to $q \rightarrow q + \gamma$ type splittings. 
Second, the inclusion of PI contributions to the cross section is seen, as expected, to lead to an increase in the cross section relative to the QED corrected partons 
across all bins, as the inclusion of $x\gamma(x,Q^2)$ opens up a new channel for lepton pair production, unaccounted for in pure QCD calculations. Since the magnitude of the photon PDF is seen to become larger at low $x$, particularly at high scales ($Q^2 = 10^4 \sim 10^8$ GeV\textsuperscript{2}) 
and $\eta \simeq \frac{1}{2}\ln{(x_1/x_2)}$ where $1$ and $2$ denote the incoming photons, the predominance of the photon at low $x$ manifests as an enhanced cross section contribution in the lower and intermediate $\eta$ bins, an effect seen to hold across all mass bins.
At high rapidities the smallness of the large-$x$ photon makes this photon contribution
smaller than the decrease due to the quark suppression noted above.  

At high $\eta$, however, the change due to QED effects in the evolution 
is seen to be of comparable in magnitude to that of PI contributions. 
 In particular, we wish to highlight that for precision calculations of electroweak effects, one requires that all the partons be consistently treated (i.e. to contain all QED splittings for the quarks and gluons in an interdependent and coupled fashion) with QED in the evolution, as well as including the photon for a consistent treatment. This is especially noteworthy since the general trend of the partons after refitting with QED has an opposing effect on the cross section compared to that of PI contributions (due to a reduction of the total quark singlet), and as such, neglecting them can in principle lead to an over-estimation of the cross section where PI contributions are simply added on top of the standard QCD result, without the compensating effect in the other partons.

In fact, at high $x,\eta$, where PI contributions are less relatively important as $x\gamma(x,Q^2)$ rapidly diminishes, the effect of refitting the partons with QED is such that even the inclusion of PI contributions after accounting for QED in the evolution leads to a cross section less than that of the standard NNLO QCD prediction. In other words, the reduction of the total quark singlet content has a greater impact than the additional cross section contributions that are available from PI processes.

\subsection{Including the ATLAS data in the global fit}

\begin{figure}
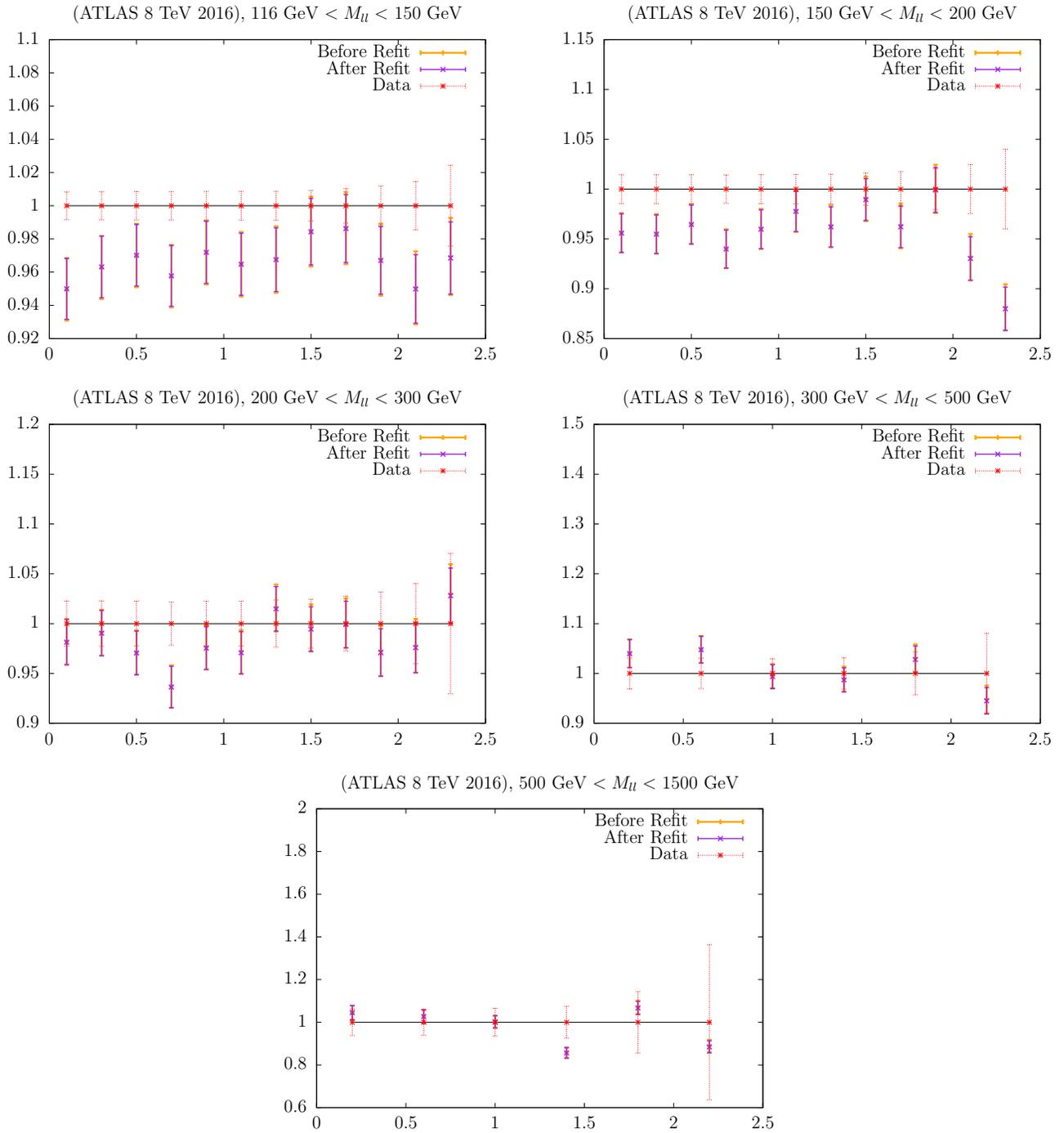

\centerline{%
\scalebox{0.7}{\input{theory_0_data_rat_noHMDY_fixed.tex}}
\scalebox{0.7}{\input{theory_1_data_rat_noHMDY_fixed.tex}}}
\centerline{\scalebox{0.7}{\input{theory_2_data_rat_noHMDY_fixed.tex}}
\scalebox{0.7}{\input{theory_3_data_rat_noHMDY_fixed.tex}}}
\centerline{\scalebox{0.7}{\input{theory_4_data_rat_noHMDY_fixed.tex}}}
\caption{The theory prediction/Data Ratio for ATLAS 8~TeV Drell-Yan data 
as a function of rapidity in different mass bins. Shown are the predictions 
using a fit in which the Drell-Yan data are not included (before refit) 
and once the Drell-Yan data are included (after refit).}
\label{fig:DY_refit_0}
\end{figure}

In the aforementioned analysis, the cross section calculations are performed using a set of PDFs 
which has not included the Drell-Yan data from ATLAS itself in the global fit for the determination of parton parameters. In the remainder of this section, we discuss the effects of including these data in the fit itself and the subsequent effect on the recalculation of the cross section. In Fig.~\ref{fig:DY_refit_0} we present the ratio of the cross section calculation from the QED corrected partons, including the contributions of PI processes, both before and after refitting to the data with these effects. We can see that there is no substantial improvement in data description after refitting.

Of note however, is the fact that the PDF contributions to the uncertainties of the predicted cross sections (the sole contribution to the uncertainty bands in Fig.~\ref{fig:DY_refit_0}) are incrementally reduced when refitting with the effects of QED included. This is best observed in the bins for high $\eta$, especially in the lower mass bins. In particular, we note that this incremental reduction is seen when refitting with the effects of QED in the evolution and with the inclusion of PI effects, but not when refit with purely with NNLO QCD parton evolution (and with no PI contributions). This indicates a weak preference to the effects of QED in the partons themselves and more accurate data may yet provide a better indication of how sensitive the comparison to the theory is with and without the effects outlined in this paper.

\section{Conclusions}

In this paper, we have presented the updated MMHT partons, modified to include the effects of QED in their evolution. Our resultant photon PDF, $x\gamma(x,Q^2)$, based on a similar methodology for the input to that of LUXqed is seen to closely resemble others in the literature, despite several modifications made to take into account our lower starting scale for the evolution and the fact that we use our own PDFs.

We have also outlined the procedure developed to provide an approximate QED corrected DGLAP evolution for the PDFs of the neutron, leading to a neutron photon PDF and isospin violating valence quark PDFs, which may hold significance for the future development of neutron PDFs. The photon PDF of the neutron is seen to be of a similar magnitude to that of the proton at higher $Q^2$.
We provide the PDFs in 
grids which contain the central sets and uncertainties. PDF sets are 
provided as grids in the LHAPDF6 format. Details are contained in the Appendix.

Finally, although the fit quality remains broadly unchanged after refitting with these effects, we have observed that for the process of high-mass Drell-Yan production, the effects of both photon initiated processes, as well as changes in the quark and antiquark PDFs due to the effects of evolution, may become significant with the advent of precision measurements in this kinematic region and that the effects of QED in the evolution may be as significant as that of the photon, highlighting a need for a fully consistent set of QED corrected partons.

\section*{Acknowledgements}

We would like to thank Valerio Bertone and Francesco Giuli.
LHL thanks the Science and Technology Facilities Council (STFC) for support via grant awards ST/P004547/1.  RST thanks the Science and Technology Facilities Council (STFC) for support via grant award ST/P000274/1. RN thanks the Elizabeth Spreadbury Fund. 

\bigskip

\section*{Appendix -- PDF Grids}\label{sec:app}

As noted earlier, the set of QED corrected partons, MMHT2015qed, developed in 
this paper will be released in the LHAPDF6 format for public use. The 
exact nature in which the grids are provided is clarified here along with
the numbering of the grids and their associated uncertainties.

The LHAPDF6 format requires that in each file for a given grid, each column, 
which represents a given PDF distribution, be labelled with an associated 
number from the Monte Carlo Particle Numbering Scheme as described in 
\cite{PDG}, where every flavour of particle is associated with an integer. 
This represents an obstacle for the photon distributions as represented in 
this paper, since only one such number is allocated for the $\gamma$, $22$, 
while we wish to distinguish between the total, the elastic and the inelastic 
components. 

To provide users with the ability to call upon $\gamma$, $\gamma^{(el)}$, 
$\gamma^{(inel)}$, as needed, we provide three separate PDF sets for each use 
case. Each set contains the full 62 eigenvector uncertainties as well as the 
central values described in  Section~\ref{ssec_unc},. The `MMHT2015qed\_nnlo\_total' 
set provides the full $\gamma = \gamma^{(el)} + \gamma^{(inel)}$ distribution 
in the column reserved for the photon (22). The `MMHT2015qed\_nnlo\_inelastic' set
provides the $\gamma^{(inel)}(x,Q^2)$ distribution while the 
`MMHT2015qed\_nnlo\_elastic' set provides the $\gamma^{(el)}(x,Q^2)$ distribution.
Users should therefore distinguish by name the appropriate LHAPDF6 variables 
in code for each distinct photon component as needed, calling each from the 
sets as labelled above.

Each grid is a file labelled as, at NLO, `mmht2015qed\_nlo\_\{type\}\_00$\{x\}$.dat' or, at NNLO,
`mmht2015qed\_nnlo\_\{type\}\_00$\{x\}$.dat', where \{type\} is a label denoting which photon contribution is included in the set and $\{x\}$ 
represents numbers in the range $\{01,02,...,62\}$. The particular 
uncertainties (as described above) associated with the numbers denoting each 
set are detailed in the Table \ref{tab:uncertainty}. 

\begin{table}[]
\begin{tabular}{|l|p{9cm}|}
\hline
\textbf{File number index $\{x\}$} & \textbf{Corresponding Uncertainty} \\ \hline
01-50 & The standard PDF uncertainties associated with the $q+\bar{q}$, $q-\bar{q}$ and $g$ distributions for all flavours \\ \hline
51-52 & The uncertainty contributions from $A'_2$ (51: -0.4, 52:-0.2) \\ \hline
53-54 & The uncertainty contributions from the Continuum contributions (53: Upper band, 54: Lower band) \\ \hline
55-56 & The uncertainty contributions from the Resonance contributions (53: Upper band, 54: Lower band) \\ \hline
57-58 & The uncertainty contributions from $W_{cut}$ (57: 3 GeV\textsuperscript{2}, 58: 4 GeV\textsuperscript{2}) \\ \hline
59-60 & The uncertainty contributions from R (59: +50\%, 60: -50\%) \\ \hline
61-62 & The uncertainty contributions from the Elastic contributions (53: Upper band, 54: Lower band) \\ \hline
\end{tabular}
\caption{A table denoting how the numbering of the grid files (produced in the LHAPDF6 format) corresponds to the uncertainties listed in the text. }
\label{tab:uncertainty}
\end{table}

\pagebreak

\bibliography{references}{}
\bibliographystyle{h-physrev}

\end{document}